\documentclass[12pt]{article}
\usepackage{amsmath}
\usepackage{amssymb}
\begin{document}
\vskip 2cm
\begin{center}
{\sf {\Large Augmented Superfield Approach to Nilpotent Symmetries of the  Modified 
Version of 2D Proca Theory}}

\vskip 3.0cm

{\sf A. Shukla$^{(a,b)}$, 
S. Krishna$^{(a,c)}$,
R. P. Malik$^{(a,d)}$}\\

$^{(a)}$ {\it Physics Department, Centre of Advanced Studies,}\\
{\it Banaras Hindu University, Varanasi - 221 005, (U.P.), India}\\

$^{(b)}$ {\it Indian Institute of Science Education and 
Research, Kolkata - 741 246, India}\\

$^{(c)}$  {\it Indian Institute of Science Education and 
Research, Mohali - 140 306,  India}

$^{(d)}$ {\it DST Centre for Interdisciplinary Mathematical Sciences,}\\
{\it Faculty of Science, Banaras Hindu University, Varanasi - 221 005, India}\\
{\small {\sf {e-mails: ashukla038@gmail.com; skrishna.bhu@gmail.com;   rpmalik1995@gmail.com}}}

\end{center}

\vskip 2cm

\noindent
{\bf Abstract:} 
We derive the complete set of off-shell nilpotent  and absolutely anticommuting
Becchi-Rouet-Stora-Tyutin (BRST), anti-BRST and (anti-)co-BRST symmetry transformations for 
{\it all} the fields of the modified version of two (1+1)-dimensional (2D) 
Proca theory by exploiting the ``augmented" superfield formalism where the (dual-)horizontality 
conditions  and (dual-)gauge-invariant restrictions  are exploited {\it together}. 
We capture the (anti-)BRST and (anti-)co-BRST invariance 
of the Lagrangian density in the language of superfield approach. We also express the nilpotency 
and absolute anticommutativity of the (anti-)BRST and (anti-)co-BRST
charges within the framework of augmented superfield formalism. This exercise leads to some {\it novel}
observations which have, hitherto, not been pointed out in the literature within the framework of 
superfield approach to BRST formalism. For the sake of completeness, we also mention, very briefly, 
a unique bosonic symmetry, the ghost-scale symmetry and discrete symmetries of the theory and show
that the algebra of conserved charges provides a physical
realization of the Hodge algebra (satisfied by the de Rham cohomological operators of differential geometry).



\newpage
\section{Introduction}

One of the earliest known gauge theories (with $U(1)$ gauge symmetry) is the Abelian 1-form 
($A^{(1)} = dx^\mu\,A_\mu$, $\mu = 0, 1, 2,...D-1$)
Maxwell theory which describes the massless vector boson ($A_\mu$) with ($D-2$)
 degrees of freedom in any arbitrary D-dimensions of spacetime. 
Thus, in the physical {\it four} dimensions of spacetime, $A_\mu$ has two degrees of freedom. 
Its massive generalization is a Proca theory that describes a vector boson with three degrees of freedom in the physical four (3+1)-dimensions
of spacetime. The central goal of our present investigation is to study the two (1+1)-dimensional (2D) Stueckelberg-modified [1] version of the Proca
theory which also incorporates a pseudo-scalar field on physical and mathematical grounds [2,3]. This model is very
{\it special} because it is endowed with {\it mass} together with various kinds of {\it internal} symmetries which originate, primarily, from the gauge symmetry and its ``dual'' version.
The existence of the above symmetries renders the model to become an example
for the Hodge theory [2,3].

Recently, in a set of papers [4-6], we have demonstrated that the 
$N = 2$ supersymmetric (SUSY) quantum mechanical models also
provide a set of physical examples of Hodge theory because of their specific continuous and 
discrete symmetry transformations which provide 
the physical realizations of the de Rham cohomological operators
and Hodge duality ($*$) operation of differential geometry [7-12]. However, these SUSY models are {\it not}
gauge theories because they are not endowed with first-class constraints in the terminology of Dirac's prescription for the classification
scheme of constraints [13,14]. One of the characteristic features of these SUSY models 
is that they have {\it mass} but do {\it not} possess gauge symmetries that are primarily generated by the first-class constraints (see, e.g. [14,15]).

We have also provided the physical realizations of the de Rham cohomological operators of the 
differential geometry in the context of Abelian $p$-form ($p = 1, 2, 3$) gauge theories 
in $D = 2 p$ dimensions of spacetime within the framework of BRST formalism [16-19]. 
As a consequence, these theories are also the field theoretic models for the Hodge theory. One of the decisive
features of these theories is the observation that they
have gauge symmetry (generated by the first-class constraints)
 but they do not have {\it mass}. The modified version
 of 2D Proca theory is, thus, a very {\it special} field theory which possesses  {\it mass} as well as various kinds of
 {\it internal} symmetries and, as has turned out, it {\it also} presents a 
field theoretic model for the Hodge theory within the
framework of BRST formalism [2,3].

One of the most intuitive approaches to understand the abstract mathematical properties associated
with the (anti-)BRST symmetries is the geometrical superfield formalism (see, e.g. [20-23]) 
where the celebrated horizontality condition (HC) plays very important role as far as the derivation of 
(anti-)BRST symmetry transformations
for the gauge fields and associated (anti-)ghost fields, for a given gauge theory, is concerned. In the augmented
version [24-27] of the above superfield formalism, the HC blends together with the gauge-invariant restrictions
(GIRs) in a beautiful fashion enabling us to derive the (anti-) BRST symmetry transformations for the
gauge, (anti-)ghost and {\it matter} fields of a given
interacting gauge theory in a cohesive and consistent manner. 
The central objective of our present paper is to apply extensively the above
augmented version of the geometrical superfield formalism [24-27] 
to discuss various aspects of the modified version of 2D Proca theory
within the framework of BRST formalism.

In our present investigation, we derive the off-shell nilpotent and absolutely anticommuting (anti-)BRST and 
(anti-)co-BRST symmetry transformations by exploiting the theoretical power of augmented version of superfield
formalism. In fact, we exploit the celebrated
(dual-)horizontality conditions [(D)HCs] and (dual-)gauge invariant
restrictions [(D)GIRs] to obtain the proper (anti-)BRST and (anti-)co-BRST 
symmetry transformations for {\it all}
the fields of the modified version of 2D Proca theory. We provide the geometrical meaning to the above
nilpotent symmetry transformations in the language of translational generators along the Grassmannian 
directions of the (2, 2)-dimensional supermanifold on which our ordinary modified version of 2D Proca
theory is generalized.

Some of the key observations of our present investigation are contained in our subsections {\bf 3.3}
and {\bf 4.4} where we have expressed the (anti-)BRST and (anti-)co-BRST charges in terms of the 
superfields (obtained after the applications of (D)HCs and (D)GIRs), Grassmannian partial derivatives and
Grassmannian differentials. The off-shell nilpotency and absolute anticommutativity properties of the
(anti-)BRST and (anti-)co-BRST symmetries (and their corresponding generators) emerge very naturally
within the framework of our augmented version of  superfield formalism. We have also captured the (anti-)BRST 
and (anti-)co-BRST invariances of the Lagrangian density within the ambit of our 
augmented version of superfield approach in a very simple and straightforward manner.


The main motivating factors behind our present investigations are as follows. First, it is very important for us 
to  put  the basic ideas of our augmented version of superfield formalism on solid footing by applying
it to various interesting physical systems which are BRST invariant. Second, it is essential for us to establish
the correctness of our earlier results [3] where we have discussed  about the off-shell nilpotent 
(anti-)BRST and (anti-)co-BRST symmetry transformations for the 2D modified Proca theory. Finally, the
present endeavor is our modest step towards the main goal of applying our basic ideas to find out the 4D massive
models for the Hodge theory which might enforce the existence of fields that would turn out to be the
candidates for the dark  matter [28,29]. We have already shown the emergence and existence of the latter
(as a pseudo-scalar field with a negative kinetic term) in our study of the modified version 
of 2D Proca theory [2,3].

The contents of our present paper are organized as follows. 
In Sec. 2, we recapitulate the bare essentials of the usual Proca theory and 
discuss the gauge symmetry transformations of the Stueckelberg-modified version of it in any 
arbitrary D-dimensions of spacetime. Our Sec. 3 is devoted to the derivation of off-shell nilpotent (anti-)BRST symmetry transformations within the framework of augmented superfield formalism. 
In Sec. 4, we deal with the (anti-)co-BRST symmetry transformations
for the 2D Stueckelberg-modified Proca theory
by exploiting the augmented version of superfield appraoch. 
Our Sec. 5 describes, very briefly,
a unique bosonic symmetry, the ghost-scale symmetry
and discrete symmetries of our present theory. 
In Sec. 6, we present the algebraic structure of  all the generators of the above
continuous  symmetries and establish its connection with the 
cohomological operators of differential geometry. 
Finally, we make some concluding remarks in Sec. 7.

In our Appendices A and B, we perform some explicit computations which have 
been used in the main body of our present text.\\

{\it Essential definitions} 
\begin{enumerate}
\item On a compact 
manifold without a boundary, a set of three
operators ($d,\, \delta,\, \Delta$) define the de Rham cohomological operators  [7-12] of 
differential geometry. Here $d = dx^\mu\,\partial_\mu$
(with $d^2 = 0$) is the exterior derivative and $\delta = \pm\, *\,d\,*$ (with $\delta^2 = 0$) 
defines the co-exterior derivative where
($*$) stands for the Hodge duality operation. The Laplacian operator $\Delta = (d + \delta)^2 = \{d, \delta\}$
is defined in terms of $d$ and $\delta$ (where $\{d, \, \delta\} = d\, \delta + \delta \, d$).
\item  We have christened the extended version of the usual Bonora-Tonin superfield formalism
[20,21] as the augmented superfield formalism where, in addition to the HC, other physically relevant restrictions
(consistent with the HC) are {\it also} imposed on the superfields defined on the appropriate supermanifold.
\end{enumerate}

\noindent

\section {Preliminaries: local gauge symmetries in the modified version of Proca theory}

Let us begin with the Lagrangian density (${\cal L}_0$) of a Proca theory 
(with  a mass parameter $m$) in any arbitrary D-dimensions of spacetime. This can be expressed in
an explicit form as follows: 
\begin{eqnarray}
{\cal L}_0 = -\frac{1}{4} F_{\mu\nu}F^{\mu\nu}\,+ \frac{m^2}{2}A_\mu A^\mu.
\end{eqnarray}
Here $F_{\mu\nu} = \partial_\mu A_\nu - \partial_\nu A_\mu$ 
is derived from  the 2-form $F^{(2)} = d A^{(1)} = [(dx^\mu \wedge dx^\nu)/2!]\; F_{\mu\nu}$
where $d = dx^\mu \partial_\mu$ (with $d^2 = 0$) is the exterior derivative and the 1-form
$A^{(1)} = dx^\mu A_\mu$ defines the vector boson $A_\mu$. In physical four (3+1)-dimensions of spacetime,
the bosonic field $A_\mu$ has three degree of freedom and $m$ has the dimension of mass
in natural units (where $\hbar = c = 1$). In the massless limit (i.e. $m = 0$), we obtain the 4D
Maxwell Lagrangian density from (1) which respects the $U(1)$ gauge invariance under the transformations:
\begin{eqnarray}
A_\mu \to A_\mu \mp \frac{1}{m}\,\partial_\mu \Lambda,
\end{eqnarray}
where $\Lambda$ is the local gauge parameter. It is evident that, in the Proca theory, the gauge symmetry 
transformations (2) are lost because of the presence of mass term. In some sense, a Proca 
theory is a generalization of  Maxwell's
theory as the latter is the massless ($m = 0$) limit of the former 
(where the usual $U(1)$ gauge symmetry invariance  is respected).

By exploiting the Stueckelberg formalism, one can restore the gauge symmetry (2) for the 
original Lagrangian density (1) where the field 
$A_\mu$ is replaced by $A_\mu \mp \frac{1}{m}\,\partial_\mu \phi$. As a consequence, we obtain the 
following Stueckelberg's Lagrangian density
\begin{eqnarray}
{\cal L}_s = - \frac{1}{4} F_{\mu\nu}F^{\mu\nu}\,+ \frac{m^2}{2}A_\mu A^\mu\,
+ \frac{1}{2}\partial_{\mu}\phi\,\partial^{\mu}\phi\, \mp \,m\,A_\mu\partial^{\mu}\phi,
\end{eqnarray}
which respects the following local, continuous and infinitesimal gauge symmetry  transformations ($\delta_g$):
\begin{eqnarray}
\delta_g\, A_\mu = \partial_\mu \Lambda, \qquad\qquad \quad \delta_g \phi = \pm \,m \,\Lambda,
\end{eqnarray}
where $\phi$ is a real scalar field. The key points, at this stage, are as follows. First, by incorporating
the Stueckelberg field $\phi$, we have converted the second-class constraints of the original Lagrangian density (1) 
into the first-class variety in the terminology of Dirac's prescription for the classification scheme [13,14]. 
Second, the Lagrangian density (3) describes, in the 
physical {\it four} dimensions of spacetime,  a theory where 
the mass and gauge invariance co-exist together in a beautiful and meaningful manner.

We close this section with the following remarks. First, the gauge symmetry transformations (4) are 
valid in any arbitrary dimension of spacetime for the Stueckelberg-modified Lagrangian density 
(${\cal L}_s$) at the {\it classical} level. This symmetry, therefore, could be exploited for 
the (anti-)BRST symmetry transformations at the {\it quantum} level. Second, the quantity 
$A_\mu \mp \frac{1}{m}\, \partial_\mu \phi$ is a gauge-invariant quantity because
$\delta_g\,[A_\mu \mp\, \frac{1}{m}\, \partial_\mu \phi] = 0$ (for $\delta_g\, A_\mu = \partial_\mu \Lambda$ 
and $\delta_g \phi = \pm \,m \Lambda$). These observations would play very important roles in our further discussions 
on the derivation of proper (anti-)BRST symmetries within the framework of augmented version of superfield formalism.

\section {Nilpotent (anti-)BRST  symmetries: geometrical  superfield  formalism}

In this section, we derive the full set of {\it proper} (anti-)BRST symmetry transformations by exploiting 
the strength of HC and GIR. Furthermore, we capture the (anti-)BRST invariance of the Lagrangian density 
and the nilpotency as well as absolute anticommutativity properties  of the (anti-)BRST charges within 
the framework of superfield formalism

\subsection{Derivation of the (anti-)BRST symmetries:  HC and GIR}

According to the prescription, laid down by the superfield approach to BRST formalism [20,21], we have to generalize
the present  D-dimensional Stueckelberg-modified theory onto a 
(D, 2)-dimensional supermanifold which is parameterized by the superspace 
variable $Z^M = (x^\mu, \theta, \bar\theta)$ where $x^\mu (\mu = 0, 1, 2,...D-1)$ are the ordinary D-dimensional spacetime variables
and ($\theta, \bar\theta$) are a pair of Grassmannian variables (with $\theta^2 = {\bar\theta}^2$ $ = 0, \; 
\theta\,\bar\theta + \bar\theta\,\theta = 0$).

The central role, in the superfield approach [20-23], is played by the HC which 
requires that the gauge-invariant quantity
$F_{\mu\nu}$, owing its origin to the exterior derivative, must remain independent of the Grassmannian 
variables when it is generalized onto a 
(D, 2)-dimensional supermanifold. In other words, 
the ordinary curvature 2-form $F^{(2)} = d A^{(1)} = (dx^\mu \wedge dx^\nu /2!) F_{\mu\nu}$ must be equal 
(i.e. $F^{(2)} = \tilde {\cal F}^{(2)}$) to the super curvature 2-form $(\tilde {\cal F}^{(2)})$:
\begin{eqnarray}
{\tilde {\cal F}}^{(2)} = \tilde d \, \tilde A^{(1)} 
\equiv  \Big(\frac {dZ^M \wedge dZ^N}{2!}\Big)\,{\tilde {\cal F}}_{MN}\,(x, \theta, \bar\theta).
\end{eqnarray}
In the above, the super exterior derivative $\tilde d$ (with ${\tilde d}^2 = 0$) and super 1-form connection ${\tilde A}^{(1)}$ are defined on the
(D, 2)-dimensional supermanifold as
\begin{eqnarray}
&& \tilde d = dZ^M\, \partial_M \equiv dx^\mu\,\partial_\mu + d\theta\, \partial_{\theta} + d\bar\theta\,\partial_{\bar\theta}, \nonumber\\
&& {\tilde A}^{(1)} = dZ^M\, A_M \equiv dx^\mu\,{\cal B}_\mu \,(x, \theta, \bar\theta) + d\theta\, \bar F\,(x, \theta, \bar\theta) 
+ d\bar\theta\,F\,(x, \theta, \bar\theta).
\end{eqnarray}
We have taken  $\partial_M = (\partial_\mu, \partial_{\theta}, \partial_{\bar\theta})$ as the 
superspace derivative on the (D, 2)-dimensional supermanifold.  Physically, the equality 
$d A^{(1)} = \tilde d \tilde A^{(1)}$ of the HC implies that the gauge-invariant electric and magnetic
fields of the ordinary theory  should {\it not} be affected by the presence of the Grassmannian  
variables $\theta$ and $\bar\theta$ of the supermanifold on which the ordinary theory has been
generalized within the ambit of superfield formalism.

The superfields ${\cal B}_\mu (x, \theta, \bar\theta),\, F (x, \theta, \bar\theta), \,\bar F(x, \theta, \bar\theta)$ 
of (6) are the generalizations of the gauge field ($A_\mu$), ghost field ($C$) and anti-ghost field ($\bar C$), respectively, of the ordinary 
D-dimensional BRST-invariant theory because the above superfields can be expanded along the Grassmannian directions of
the (D, 2)-dimensional supermanifold as (see, e.g. [20])
\begin{eqnarray}
{\cal B}_\mu\, (x, \theta, \bar\theta) &=& A_\mu(x) + \theta\,R^{(1)}_\mu (x) +\bar\theta\, R^{(2)}_\mu (x) 
+ i\,\theta\bar\theta\, S_\mu (x),\nonumber\\
F\,(x, \theta, \bar\theta) &=& C(x) + i\,\theta\,B_1 (x)  + i\,\bar\theta\, B_2 (x)  + i\,\theta\bar\theta\, s(x), \nonumber\\
\bar F \,(x, \theta, \bar\theta) &=& \bar C (x) + i\,\theta\,B_3 (x)  + \bar\theta\, B_4 (x)  + i\,\theta\bar\theta\, \bar s(x),
\end{eqnarray}
where ($A_\mu, C, \bar C$) are the basic fields of any arbitrary D-dimensional
(anti-)BRST invariant Abelian theory\footnote{An Abelian BRST invariant theory, in any
arbitray dimension of spacetime,  contains the gauge-fixing
 and FP-ghost terms in addition to the kinetic term for $A_\mu$.} and rest of the fields, on  the r.h.s.
of (7), are the secondary fields  which can be expressed in terms of the basic and auxiliary fields of the ordinary
D-dimensional theory by exploiting HC. It is clear that ($ R^{(1)}_\mu,\, R^{(2)}_\mu, s, \, \bar s$) and 
($S_\mu, \, B_1,\, B_2,\, B_3,\, B_4$) are the fermionic and bosonic secondary fields, respectively, on the r.h.s. of (7).

One can expand the expression $\tilde d \, \tilde A^{(1)}$ of (5) in the following explicit form using (6).
This expansion, in its full blaze of glory, is as follows
\begin{eqnarray}
\tilde d\, \tilde A^{(1)} &=& \left( \frac{dx^\mu \wedge dx^\nu}{2!}\right)\,(\partial_\mu {\cal B}_\nu 
- \partial_\nu {\cal B}_\mu ) + (dx^\mu \wedge d\theta)\,(\partial_\mu \bar F 
- \partial_{\theta} {\cal B}_\mu)   \nonumber\\ 
&+&  (dx^\mu \wedge d\bar\theta)\,(\partial_\mu F - \partial_{\bar\theta} {\cal B}_\mu) 
- (d\theta \wedge d\theta)(\partial_\theta \bar F)
 - (d\bar\theta \wedge d\bar\theta)(\partial_{\bar\theta} F) \nonumber\\ 
&-& (d\theta \wedge d\bar\theta) (\partial_\theta F + \partial_{\bar\theta} \bar F).
\end{eqnarray}
In a similar fashion, one can also expand the r.h.s. of (5) as follows:
\begin{eqnarray}
&& \Big(\frac{dx^\mu \wedge dx^\nu}{2!}\Big)\,\tilde {\cal F}_{\mu\nu} +
\left(dx^\mu \wedge d\theta \right)\,\tilde {\cal F}_{\mu\theta} 
+ \left(dx^\mu \wedge d\bar\theta \right)\,\tilde {\cal F}_{\mu\bar\theta}
\nonumber\\ &&+ \Big(\frac{d\theta \wedge d\theta}{2!}\Big)\,\tilde {\cal F}_{\theta\theta}
+ \Big(\frac{d\bar\theta \wedge d\bar\theta}{2!}\Big)\,\tilde {\cal F}_{\bar\theta\bar\theta}
+ \left( d\theta \wedge d\bar\theta\right)\tilde {\cal F}_{\theta\bar\theta}.
\end{eqnarray}
The  HC requires that $F^{(2)} = [dx^\mu \wedge dx^\nu/2!]\, F_{\mu\nu}$ should be equal to 
$\tilde {\cal F}^{(2)} = [d Z^M \wedge d Z^N/2!]\, {\cal F}_{MN} $. This implies that 
$\tilde {\cal F}_{\mu\theta} = \tilde {\cal F}_{\mu \bar\theta} = \tilde {\cal F}_{\theta\theta}
 = \tilde {\cal F}_{\theta \bar\theta} = \tilde{\cal F}_{\bar\theta \bar\theta} = 0$.

Written in an explicit form, we have the following 
relationships (from the comparison between  (8) and (9)) due to the celebrated HC, namely;
\begin{eqnarray}
&&\tilde {\cal F}_{\mu\theta} = \partial_\mu \bar F - \partial_{\theta} {\cal B}_\mu, 
\quad \tilde {\cal F}_{\mu\bar\theta} = \partial_\mu F - \partial_{\bar\theta} {\cal B}_\mu, \quad
\frac{1}{2!} \; \tilde {\cal F}_{\theta\theta} = - \partial_\theta \bar F,  \quad
\frac{1}{2!} \; \tilde {\cal F}_{\bar\theta\bar\theta} = - \partial_{\bar\theta} F, \nonumber\\
&&\tilde {\cal F}_{\theta\bar\theta} = -(\partial_\theta F + \partial_{\bar\theta} \bar F) \qquad
\tilde {\cal F}_{\mu\nu} \equiv (\partial_\mu {\cal B}_\nu - \partial_\nu {\cal B}_\mu ) \Longrightarrow 
 F_{\mu\nu} \equiv (\partial_\mu A_\nu - \partial_\nu A_\mu). 
\end{eqnarray}
The substitution of the expansions (7) into the above equation  yields the following relationships amongst  
the secondary fields and the basic as well as auxiliary fields of the 
ordinary 2D theory\footnote{Our method of derivation is somewhat different 
from the original work of Bonora-Tonin superfield formalism [20,21] even though our present work is motivated by
the latter works (i.e. [20,21]).}, namely;
\begin{eqnarray}
&&R^{(2)}_\mu = \partial_\mu C, \quad R^{(1)}_\mu = \partial_\mu \bar C, \quad \,\,S_\mu = \partial_\mu B_4 \equiv - \partial_\mu B_1, \nonumber\\
&&B_2 = B_3 = 0, \qquad s = \bar s = 0, \qquad  B_4 + B_1 = 0.
\end{eqnarray}
The last entry in the above is nothing but the celebrated Curci-Ferrari condition [30] which turns out to be trivial
in the case of  Abelian 1-form modified  Proca gauge theory.
Taking the help of (11), we have the following expansions (if we choose $B_4 = B,  B_1 = - B$), namely;
\begin{eqnarray}
{\cal B}^{(h)}_\mu\, (x, \theta, \bar\theta) &=& A_\mu(x) + \theta\;(\partial_\mu \bar C) 
+\bar\theta\; (\partial_\mu C) 
+ \theta\bar\theta\; (i\, \partial_\mu B)\nonumber\\
&\equiv & A_\mu(x) + \theta\;(s_{ab}\; A_\mu ) + \bar\theta\;(s_b \; A_\mu ) 
+ \theta\bar\theta\; (s_b\,s_{ab}\; A_\mu ), \nonumber\\
F^{(h)}\,(x, \theta, \bar\theta) &=& C(x) + \theta\; (-i\,B) + \bar\theta\;(0)   + \theta\bar\theta \;(0) \nonumber\\
&\equiv & C + \theta\;(s_{ab}\; C ) + \bar\theta\;(s_b \; C ) + \theta\bar\theta \; (s_b\,s_{ab}\; C ), \nonumber\\
\bar F^{(h)}\,(x, \theta, \bar\theta) &=& \bar C (x) + \theta\;(0) + \bar\theta\;(i\, B)  + \theta\bar\theta\;(0) \nonumber\\
&\equiv & \bar C + \theta\;(s_{ab}\; \bar C ) + \bar\theta\;(s_b \; \bar C ) 
+ \theta\bar\theta\; (s_b\,s_{ab}\; \bar C ),
\end{eqnarray}
which yields the following off-shell nilpotent (anti-)BRST symmetries for the gauge ($A_\mu$) 
and Faddeev-Popov (FP) (anti-)ghost fields $(\bar C)C$ of the theory:
\begin{eqnarray}
&& s_b\, A_\mu = \partial_\mu C,\;\;\qquad s_b \,C = 0,\;\;\qquad s_b\, \bar C =i\,B,\;\;\qquad s_b\, B = 0, \nonumber\\
&& s_{ab} \, A_\mu = \partial_\mu \bar C, \qquad s_{ab}\, \bar C = 0, \qquad s_{ab} \,C = -i\,B, \qquad s_{ab} \,B = 0.
\end{eqnarray}
A few noteworthy points, at this stage, are as follows. First, the superscript ($h$) on the 
superfields  in (12) denotes the expansion of the superfields after the application of HC. 
Second, the transformations $s_{(a)b}\, B = 0$ on the Nakanishi-Lautrup auxiliary fields 
$B$ have been derived from the nilpotency condition. Third, it can be checked that the last entry of (10) is satisfied: 
$\tilde {\cal F}^{(h)}_{\mu\nu} = \partial_\mu {\cal B}^{(h)}_\nu - \partial_\nu {\cal B}^{(h)}_\mu \, \equiv \,
\partial_\mu A_\nu - \partial_\nu A_\mu = F_{\mu\nu}$ due to expansions in (12).
Finally, we have the following mappings  (see, e.g. [24-27] for details)
\begin{eqnarray}
&& s_b\,\longleftrightarrow \,\lim_{\theta = 0} \frac {\partial}{\partial{\bar\theta}},  \quad\qquad 
s_{ab}\,\longleftrightarrow \,\lim_{\bar\theta = 0} \frac {\partial}{\partial{\theta}},\quad \qquad 
s_b\, s_{ab}\,\longleftrightarrow \,\frac{\partial}{\partial \bar\theta} \,\frac {\partial}{\partial{\theta}}.
\end{eqnarray}
Thus, we note that the Grassmannian translation generators ($\partial_\theta,\, \partial_{\bar\theta}$) 
are connected with the off-shell nilpotent ($s^2_{(a)b} = 0$) and absolutely anticommuting 
($s_b\,s_{ab} + s_{ab}\, s_b = 0$) fermionic (anti-)BRST symmetry transformations $s_{(a)b}$.
These properties have their origin in the properties $\partial^2_\theta = 0, \,\partial^2_\theta = 0,\, 
\partial_\theta\partial_{\bar\theta} + \partial_{\bar\theta}\partial_\theta = 0$ of the
Grassmannian translation generators ($\partial_\theta, \partial_{\bar\theta}$) when the above relations
are taken in their operator form.

Truly speaking, the exact relationship between the (anti-)BRST symmetry transformations $s_{(a)b}$ and Grassmannian translational 
generators ($\partial_\theta, \partial_{\bar\theta}$) are: 
$s_b\, M(x)$ $ = \big[ \frac{\partial}{\partial\bar\theta} {\tilde M}^{(h)} (x, \theta, \bar\theta)\big]| _{\theta = 0}$ for the BRST transformations
and $s_{ab}\, M(x) = \big[ \frac{\partial}{\partial\theta} 
{\tilde M}^{(h)}\, (x, $ $ \theta, \bar\theta)\big]| _{\bar\theta = 0}$ for the anti-BRST symmetry tranformations
where $M(x)$ is the D-dimensional ordinary field and $\tilde M^{(h)}(x, \theta, \bar\theta)$ is the corresponding superfield (obtained after the application of the HC). However, we shall continue with the mapping (14) but shall keep in mind that
the precise connection between the (anti-)BRST transformations $s_{(a)b}$ and 
$(\partial_\theta, \partial_{\bar\theta})$ is:
 $s_b  \leftrightarrow  \partial_{\bar\theta} $ and $s_{ab} \leftrightarrow  \partial_{\theta}$.

Now we exploit the strength of the augmented version of superfield formalism [24-27] to derive the (anti-)BRST
symmetry transformations for the real scalar field $\phi$. To this end in mind, we recall that the quantity 
($A_\mu \mp\, \frac{1}{m}\, \partial_\mu \phi$) is a {\it gauge invariant} quantity (cf.  Sec. 2). Thus, this physical quantity
should remain unaffected by the presence of the Grassmannian variables ($\theta,\, \bar\theta$) when it is generalized
onto a (D, 2)-dimensional 
supermanifold. In other words, in the language of differential geometry, the following is true:
\begin{eqnarray}
&&\tilde d\, \tilde A_{(h)}^{(1)} (x, \theta, \bar\theta)
\; \mp\; \frac{1}{m} \;\tilde d\, \Phi (x, \theta, \bar\theta) 
= \;d\,A^{(1)} (x) \;\mp\; \frac{1}{m}\, d\, \phi (x).
\end{eqnarray}
Here $\tilde A_{(h)}^{(1)} = dx^\mu\, {\cal B}^{(h)}_\mu + d\theta\, \bar F^{(h)} + d\bar\theta\, F^{(h)}$
[cf. (12)] and the zero-form superfield $\Phi(x, \theta, \bar\theta)$ has the following super-expansion along the 
Grassmannian directions ($\theta, \bar\theta$) of the (D, 2)-dimensional supermanifold, namely;
\begin{eqnarray}
\Phi (x, \theta, \bar\theta) = \phi(x) + \theta\; \bar f(x) + \bar\theta\; f(x) + i\, \theta\bar\theta\; b(x).
\end{eqnarray}
In the above, it is evident that the pair 
of secondary fields ($f(x),\, \bar f(x)$) are fermionic and ($\phi(x),\, b(x)$)
are bosonic in nature. In the limit $(\theta,\, \bar\theta) = 0$, 
we retrieve back our real scalar Stueckelberg field $\phi(x)$ of the original D-dimensional ordinary theory.

The gauge-invariant restriction (GIR) in (15) leads to the following relationships:
\begin{eqnarray}
 \bar f = \pm\, m\, \bar C , \qquad   f = \pm\, m \, C, \qquad b = \pm\, m\, B.
\end{eqnarray}
The substitution of (17) into (16) yields
\begin{eqnarray}
\Phi^{(g)} (x, \theta, \bar\theta) &=& \phi(x) + \theta\; (\pm\, m\, \bar C) + \bar\theta\;(\pm\, m \,C) 
+ \theta\bar\theta\; (\pm \,i\, m\, B), \nonumber\\
&\equiv & \phi(x) + \theta\; (s_{ab} \;\phi) + \bar\theta\;(s_b\; \phi) 
+ \theta\bar\theta\; (s_b \,s_{ab}\; \phi),
\end{eqnarray} 
where the superscript ($g$) on the superfield $\Phi(x, \theta, \bar\theta)$ corresponds to the super-expansion of this superfield 
after the application of GIR. It is clear, from the above equation, that we have the following:
\begin{eqnarray}
 && s_b\, \phi = \pm\, m \,C , \qquad   s_{ab}\, \phi = \pm\, m\, \bar C, \qquad 
s_b \,s_{ab}\, \phi = \pm\, i\, m\, B.
\end{eqnarray}
We note that, ultimately, it is the combination of HC and GIR which leads to the derivation of 
 full set of correct off-shell nilpotent
($s^2_{(a)b} = 0$) and absolutely anticommuting ($s_b\, s_{ab} + s_{ab}\, s_b = 0$) (anti-)BRST 
transformations for {\it all} the fields of the D-dimensional modified version of Proca theory.

\subsection{Lagrangian densities:  (anti-)BRST   invariance}

Exploiting the full set of (anti-)BRST symmetry transformations, we can derive the (anti-) BRST invariant Lagrangian densities 
(that incorporate the gauge-fixing and Faddeev-Popov ghost terms) by exploiting the standard techniques of BRST approach, namely;
\begin{eqnarray}
 {\cal L}_B &=& {\cal L}_s + s_b\,s_{ab}\; \Big[\frac{i}{2}\, A_\mu \,A^\mu - \,\frac{i}{2}\, \phi^2  
+ \frac{1}{2}\, C \,\bar C \Big], \nonumber\\
 &=& {\cal L}_s + s_b\; \Big[i\, (A_\mu \,\partial^\mu \bar C \mp m\,\phi \,\bar C - \frac{1}{2}\, B\, \bar C) \Big ],\nonumber\\
&=& {\cal L}_s + s_{ab}\; \Big[-i\, (A_\mu\, \partial^\mu  C \mp m\,\phi \, C - \frac{1}{2}\, B\, C ) \Big ].
\end{eqnarray}
In explicit form, the total (anti-)BRST invariant Lagrangian densities
(containing two signatures) look in the following form:
\begin{eqnarray}
{\cal L}_B &=& - \frac{1}{4} F_{\mu\nu}F^{\mu\nu}\,+ \frac{m^2}{2}A_\mu A^\mu\, 
+ \frac{1}{2}\partial_{\mu}\phi\,\partial^{\mu}\phi\,  \mp  \,m\,A_\mu\partial^{\mu}\phi 
\nonumber\\ &+& B(\partial \cdot A \pm m\,\phi) 
+\frac{1}{2}\, B^2 
- i\,\partial_\mu \bar C\, \partial^\mu C + i\, m^2 \bar C\, C.
\end{eqnarray}
Using the full set of (anti-)BRST symmetries (13) and (19), we can check that the above 
Lagrangian densities transform to
the total spacetime derivatives as:  
\begin{eqnarray}
s_b\,{\cal L}_B =\partial_\mu \;\bigl [B\, \partial^\mu C \bigr ], \qquad
s_{ab}\,{\cal L}_B =\partial_\mu\; \bigl [B\, \partial^\mu \bar C \bigr ].
\end{eqnarray}
As a consequence, the action integral 
$S = \int d^{D-1} x\; {\cal L}_B$ remains invariant for the physically well-defined fields of the theory.
The above  infinitesimal and continuous symmetry transformations, according to Noether's theorem, 
lead to the following expressions for the (anti-)BRST
charges $Q_{(a)b}$:
\begin{eqnarray}
 &&Q_{ab} =\int d^{D-1}x \;[B\, \dot{\bar C} - \dot B\, \bar C], \qquad
 Q_b =\int d^{D-1}x \;[B\, \dot{C} - \dot B\, C],
\end{eqnarray}
which are found to be conserved ($\dot {Q}_{(a)b} = 0$) and nilpotent ($ Q^2_{(a)b} = 0$) of order two. 
These charges are the
generators of transformations listed in (13) and (19) and they are derived from the following
Noether conserved currents:
\begin{eqnarray}
J^\mu_b = - F^{\mu\nu}(\partial_\nu C) + B(\partial^\mu C) + m C (\partial^\mu \phi - m  A^\mu), \nonumber\\
J^\mu_{ab} = - F^{\mu\nu}(\partial_\nu \bar C) + B(\partial^\mu \bar C) + m\bar C (\partial^\mu \phi - m  A^\mu).
\end{eqnarray}
In the proof of the conservation laws ($\partial_\mu J^\mu_{(a)b} = 0$), we have to use the
following Euler-Lagrange (E-L) equations of motion 
\begin{eqnarray}
&& (\Box + m^2)\,C = 0, \quad\qquad (\Box + m^2)\,A_\mu - \partial_\mu 
\,(\partial \cdot A \pm m \, \phi + B) = 0,\nonumber\\   
&&(\Box + m^2)\, \bar C = 0, \quad \Box\, \phi - m\,(\partial \cdot A + B) = 0, 
\quad B = -(\partial \cdot A \pm m\, \phi),
\end{eqnarray}
that emerge from the Lagrangian densities (21).

The (anti-)BRST invariance of the Lagrangian density (21) can be also captured in the 
language of superfield formalism.
To this end in mind, first of all, we note that the Stueckelberg 
Lagrangian density ${\cal L}_s$ [cf. (3)] can be written as  
\begin{eqnarray}
{\cal L}_s \rightarrow  \tilde {\cal L}_s &=& - \frac{1}{4}\, \tilde{\cal F}^{(h)}_{\mu\nu} 
\tilde{\cal F}^{\mu\nu(h)}\,
+ \frac{m^2}{2}\, {\cal B}^{(h)}_\mu {\cal B}^{\mu(h)} \nonumber\\  
&+& \frac{1}{2} \,\partial_{\mu}\Phi^{(g)}\,\partial^{\mu}\Phi^{(g)}\, 
\mp \,m\, {\cal B}^{(h)}_\mu\partial^{\mu} \Phi^{(g)},
\end{eqnarray}
within the framework of superfield formalism where the superfields are obtained after the 
applications of HC and GIR [cf. (12),(18)]. 
It is straightforward to check that the following is true, namely;
\begin{eqnarray}
\lim_{\theta = 0}\, \frac{\partial}{\partial\bar\theta}\; \tilde {\cal L}_s = 0, \quad \qquad
\lim_{\bar\theta = 0}\, \frac{\partial}{\partial\theta}\; \tilde {\cal L}_s = 0,  \quad \qquad
\frac{\partial}{\partial\bar\theta}\,\frac{\partial}{\partial\theta}\; \tilde {\cal L}_s = 0.
\end{eqnarray}
In view of the mappings (14), it is evident that the Stueckelberg Lagrangian densities are (anti-)BRST invariant
(i.e. $s_b\, {\cal L}_s = 0, \, s_{ab}\, {\cal L}_s = 0,\,s_b\,s_{ab}\, {\cal L}_s = 0$).

Exploiting the techniques  of superfield formalism, the full 
(anti-)BRST invariant Lagrangian densities (21) (that incorporate the gauge-fixing 
and Faddev-Popov ghost terms) can be expressed in {\it three} different ways
(modulo a total spacetime derivative), namely;
\begin{eqnarray}
\tilde {\cal L}_B &=& \tilde {\cal L}_s + \frac{\partial}{\partial\bar\theta}\,\frac{\partial}{\partial\theta}
\Big[\frac{i}{2}\, {\cal B}^{(h)}_\mu \,{\cal B}^{\mu(h)} - \,\frac{i}{2}\, (\Phi^{(g)}\,\Phi^{(g)}) 
+ \frac{1}{2}\, (F^{(h)}\, \bar F^{(h)}) \Big], \nonumber\\
&\equiv& \tilde {\cal L}_s + \lim_{\theta = 0} \,\frac{\partial}{\partial\bar\theta}\Big[i\, \Big\{ {\cal B}^{(h)}_\mu \,\partial^\mu \bar F^{(h)}
\mp \, m\, (\Phi^{(g)}\,\bar F ^{(h)}) 
- \frac{1}{2}\, (B (x)\, \bar F^{(h)}) \Big\} \Big], \nonumber\\
&\equiv& \tilde {\cal L}_s + \lim_{\bar\theta = 0} \frac{\partial}{\partial\theta}\Big[-i \Big\{{\cal B}^{(h)}_\mu \partial^\mu F^{(h)}
\mp  m (\Phi^{(g)} F ^{(h)}) 
- \frac{1}{2} (B (x) F^{(h)}) \Big\} \Big].
\end{eqnarray}
Taking into account the nilpotency ($\partial^2_{\theta} = \partial^2_{\bar\theta} = 0$) and anticommutativity 
($\partial_{\theta}\, \partial_{\bar\theta} \,+\, \partial_{\bar\theta}\, \partial_{\theta} = 0$) property of the generator 
($\partial_{\theta}, \,\partial_{\bar\theta}$), it is straightforward to prove that 
\begin{eqnarray}
&&\lim_{\bar\theta = 0}\, \frac{\partial}{\partial\theta}\; \tilde{\cal L}_B   =0\quad \longleftrightarrow \quad s_{ab}\, {\cal L}_B = 0, \nonumber\\
&&\lim_{\theta = 0}\, \frac{\partial}{\partial \bar\theta}\; \tilde{\cal L}_B = 0\quad \longleftrightarrow \quad s_b\, {\cal L}_B = 0, \nonumber\\
&&\frac{\partial}{\partial \bar\theta}\,\frac{\partial}{\partial\theta}\; \tilde{\cal L}_B = 0 \quad \longleftrightarrow \quad s_b\,s_{ab}\; {\cal L}_B = 0,
\end{eqnarray} 
where the mappings (14) and results from (27) have been taken into consideration. We would like to lay emphasis 
on the fact that there is {\it no} contradiction amongst (20), (22), (28) and (29). This is due to the observation
that, in reality, we have: $(1/2) \,s_b s_{ab} [i\, A_\mu A^\mu - i\, \phi^2 + C\, \bar C] =- \partial_\mu (A^\mu B)
+ B \,(\partial \cdot A \pm m\, \phi) + (B^2/2) - i\, \partial_\mu \bar C \partial^\mu C + i  m^2 \bar C C $. However,
we have thrown away the total spacetime derivative term from the Lagrangian density (21). If we keep this term in
(21), then, we have $s_{(a)b} {\cal L}_B = 0$ instead of the expressions in (22).
Thus, there is no inconsistency anywhere.

\subsection{Conserved charges: superfield approach}

We can also express the (anti-)BRST charges in terms of superfields (obtained after the application of HC and GIR), 
the Grassmannian partial derivatives ($\partial_{\theta},\, \partial_{\bar\theta}$) and differentials ($d\theta,\, d\bar\theta$).
For instance, we note that the following expression for the BRST charge is true, namely;
\begin{eqnarray}
Q_b = \frac{\partial}{\partial \bar\theta}\, \frac{\partial}{\partial\theta}\; 
\Bigl(\int d^{D-1}x\; \Big[ i\, F^{(h)}\,
 {{\cal B}}^{(h)}_0 \Big] \Bigr) 
 = \int d^{D-1}x \;\int d\bar\theta\,\int d\theta \;\Big[ i\, 
F^{(h)}\, {{\cal B}}^{(h)}_0 \Big] ,
\end{eqnarray}
in the language of superfields (after the application of HC). 
It is clear, from the mappings (14), that the above expression implies:
\begin{eqnarray}
Q_b =\int d^{D-1}x \;\Big[ s_b\, s_{ab}\,\Big( i\, C\, {A_0} \Big) \Big],
\end{eqnarray}
in the ordinary D-dimensions of spacetime where the (anti-)BRST transformations
$s_{(a)b}$ and ordinary fields are defined. 
The proof of the nilpotency of the BRST charge becomes quite  simple now due to the nilpotency ($s^2_b = 0$) of $s_b$ 
and that of the translation generator $\partial_{\bar\theta}$ (because $\partial^2_{\bar\theta} = 0$).
In exactly similar fashion, we can express the anti-BRST charge $Q_{ab}$,
within the framework of superfield formalism, as:
\begin{eqnarray}
Q_{ab} &=& \frac{\partial}{\partial\theta}\, \frac{\partial}{\partial\bar\theta}
 \; \Big(\int d^{D-1}x \Big[ -i\, {\bar F}^{(h)}\,
 {{\cal B}}^{(h)}_0 \Big] \Big) \nonumber\\
& = &\int d^{D-1}x\; \int d\theta\,\int d\bar\theta\; \Big[- i\,{\bar F}^{(h)}\, {{\cal B}}^{(h)}_0 \Big].
\end{eqnarray}
Once again, the proof of nilpotency of the anti-BRST charge $Q_{ab}$  becomes pretty simple because of the fact 
that, in the ordinary D-dimensional spacetime, the expression (32) can be written in the following form
by exploiting the mappings (14), namely;
\begin{eqnarray}
Q_{ab} =\int d^{D-1}x\; \Big[ s_{ab}\, s_b\;\Big( -i\, \bar C\, {A_0} \Big) \Big],
\end{eqnarray}
where $s^2_{ab} = 0$ implies that $Q^2_{ab} = 0$ (due to $s_{ab}\,Q_{ab} = i\, \{Q_{ab}, Q_{ab}\} = 0$).
Within the framework of superfield formalism, the nilpotency of $Q_{ab}$ is  encoded in the nilpotency of $\partial_{\theta}$
(because $\partial^2_\theta = 0$). In other words, we can explicitly verify the nilpotency
of the conserved (anti-)BRST charges in terms of the translational generators along the Grassmannian 
directions ($\theta, \bar\theta$) of the (2, 2)-dimensional supermanifold as: 
$\lim_{\theta = 0} \, ({\partial} / {\partial \bar\theta}) \, Q_b = 0, \, 
\lim_{\bar\theta = 0} \,({\partial} / {\partial \theta})\, Q_{ab} = 0$
because $\partial^2_{\theta} = \partial^2_{\bar\theta} = 0$.

There are other alternative forms of the conserved and nilpotent (anti-)BRST charges, 
within the framework of the superfield 
formalism, that are {\it also} interesting. For instance, it can checked that the 
anti-BRST charge can be expressed as:
\begin{eqnarray}
Q_{ab} &=& \int d^{D-1}x\; \int d\theta\;\Big[B(x) \,{\cal B}^{(h)}_0 (x, \theta, \bar\theta) 
+ i\, {\bar F}^{(h)}(x, \theta, \bar\theta)\, \dot {F}^{(h)}(x, \theta, \bar\theta)\Big ] \nonumber\\
&=& \lim_{\bar\theta = 0} \,\frac{\partial}{\partial\theta}\, \int d^{D-1}x\; \Big[B(x) \,{\cal B}^{(h)}_0 (x, \theta, \bar\theta) 
+ i\, {\bar F}^{(h)}(x, \theta, \bar\theta)\, \dot F^{(h)}(x, \theta, \bar\theta)\Big]\nonumber\\
&\equiv& \int d^{D-1}x\; \bigl [ s_{ab} \, (B\, A_0 + i\, \bar C\, \dot C) \, \bigr ].
\end{eqnarray}
In exactly similar fashion, we can express the conserved BRST charge as:
\begin{eqnarray}
Q_b &=& \int d^{D-1}x\; \int d \bar\theta\;\Big[B(x)\,{\cal B}^{(h)}_0 (x, \theta, \bar\theta) 
- i\,{F}^{(h)}(x, \theta, \bar\theta)\,{\dot {\bar F}^{(h)}} (x, \theta, \bar\theta)\Big ], \nonumber\\
&=& \lim_{\theta = 0} \,\frac{\partial}{\partial\bar\theta}\, 
\int d^{D-1}x\; \Big[B(x)\;{\cal B}^{(h)}_0 (x, \theta, \bar\theta) 
- i\,{F}^{(h)}(x, \theta, \bar\theta)\, {\dot {\bar F}^{(h)}} (x, \theta, \bar\theta)\Big] \nonumber\\
&\equiv& \int d^{D-1}x\; \bigl [ s_b \, (B\, A_0 - i\, C\, \dot {\bar C}) \, \bigr ].
\end{eqnarray}
The nilpotency ($Q^2_{(a)b} = 0$) of the (anti-)BRST charges $Q_{(a)b}$ 
is encoded in the observation that the following are true, namely;
\begin{eqnarray}
\lim_{\bar\theta = 0} \,\frac{\partial}{\partial \theta}\; Q_{ab} = 0, \qquad \quad 
\lim_{\theta = 0} \,\frac{\partial}{\partial\bar \theta}\; Q_b = 0,
\end{eqnarray}
where the nilpotency ($\partial^2_\theta = \partial^2_{\bar\theta} = 0 $) of $\partial_\theta$ and $\partial_{\bar\theta}$
plays an important role.

We close this subsection with the remark that the following observations in the context of expressions for the (anti-)BRST charges:
\begin{eqnarray}
&& Q_{ab} = \int d^{D-1}x \;\Big[ s_b (-i\, \bar C\, \dot {\bar C}) \Big], \qquad
 Q_b = \int d^{D-1}x \;\Big[ s_{ab} (i\, C\, \dot C) \Big], 
\end{eqnarray}
lead to the proof of absolute anticommutativity of the (anti-)BRST 
charges because it can be readily checked that
the following are true, namely; 
\begin{eqnarray}
&&s_b\, Q_{ab} = i\, \{ Q_{ab}, Q_b\} 
= \int d^{D-1}x \;\Big[ s^2_b\; \Big(-i\, \bar C\, \dot{\bar C}\Big)\Big] = 0, \quad \Big(s^2_b = 0 \Big),\nonumber\\
&&s_{ab}\, Q_b = i\, \{ Q_b, Q_{ab}\} 
= \int d^{D-1}x \;\Big[ s^2_{ab} \;\Big(i\, C\, \dot{C}\Big)\Big] = 0, \;\quad \Big(s^2_{ab} = 0 \Big),
\end{eqnarray}
because of the nilpotency of (anti-)BRST transformations $s_{(a)b}$.
These observations can also be captured in the language of the superfield formalism, namely;
\begin{eqnarray}
&& Q_{ab} = \lim_{\theta = 0} \frac{\partial}{\partial\bar\theta}\,\int d^{D-1}x\; \Big[-i\, \bar F^{(h)}\, \dot{\bar F}^{(h)} \Big],\nonumber\\
&& Q_b = \lim_{\bar\theta = 0} \frac{\partial}{\partial\theta}\,\int d^{D-1}x \;\Big[i\,  F^{(h)}\, \dot{ F}^{(h)} \Big]. 
\end{eqnarray}
The above expressions imply the following:
\begin{eqnarray}
\lim_{\theta  = 0}\, \frac{\partial}{\partial\bar\theta}\; Q_{ab} = 0,\qquad \qquad 
\lim_{\bar\theta = 0}\,\frac{\partial}{\partial\theta}\; Q_b = 0.
\end{eqnarray}
A close look at (38), (39) and (40) shows that the nilpotency and 
anticommutativity property are inter-related. In other words,
the properties  $\partial^2_\theta = \partial^2_{\bar\theta} = 0$ and $\partial_\theta \,\partial_{\bar\theta}  
+ \partial_{\bar\theta}\,\partial_{\theta} = 0$ are inter-connected. For instance, in the latter relation when we set
$\partial_\theta = \partial_{\bar\theta}$, we obtain the former relation $\partial^2_{\theta} = \partial^2_{\bar\theta} = 0$
which actually provides the connection between the properties of
anticommutativity and nilpotency associated with the (anti-)BRST symmetry transformations ($s_{(a)b}$).

We wish to make a final  remark that it  is the strength of the superfield approach
to BRST formalism that we have obtained various expressions for the (anti-)BRST charges in the language of
(anti-)BRST symmetry transformations. Some of the results are completely {\it novel} as, to the best of our knowledge,
these expressions have not been pointed out in the literature. In fact, these new expressions are responsible,
with the help of mapping in (14), to establish the nilpotency and absolute anticommutativity
of the (anti-)BRST symmetries (and corresponding charges) within the framework
of superfield formalism.  For instance, the relationships, given in (38), demonstrate that the nilpotency property
and absolute anticommutativity property (of $s_{(a)b}$ and $Q_{(a)b}$) are intertwined
together.

\section{(Anti-)co-BRST symmetries: superfield  approach}

In this section, first of all, we discuss about the dual-gauge transformations for the gauge-fixed Lagrangian
densities and show that a particular kind of restriction must be imposed on the dual-gauge
parameter if we wish to maintain the dual-gauge symmetry in the theory. Then, we
derive the off-shell nilpotent and absolutely anticommuting (anti-)co-BRST symmetry
transformations by exploiting the strength of dual-HC (DHC) and dual-GIR (DGIR). 
After this, we prove the (anti-)co-BRST invariance of the Lagrangian densities within the framework of 
superfield formalism. Finally, we capture the (anti-)co-BRST
invariance of the conserved charges, their nilpotency and absolute anticommutativity 
within the  ambit of the augmented version of superfield approach to BRST formalism.

\subsection{Dual-gauge transformations for the gauge-fixed  Lagrangian densities in two-dimensions 
 of spacetime}

Analogous to the infinitesimal, continuous and local gauge symmetry transformations (4), 
we wish to discuss, in this subsection, the
dual-gauge transformations which would be, finally, generalized to the (anti-)co-BRST 
symmetry transformations\footnote{In the two (1 + 1)-dimensions of spacetime, 
a particular part of the Lagrangian density 
[i.e. $- (1/4)\, F^{\mu\nu}\, F_{\mu\nu}$] becomes [$(1/2)\, E^2$] because 
there is only one non-vanishing  component of $F_{\mu\nu} = \partial_\mu A_\nu - \partial_\nu\, A_\mu$ which is
$F_{01} = - \varepsilon^{\mu\nu} \partial_\mu A_\nu = E$. This is a pseudo-scalar field  
because it changes sign under the operation of  parity transformation
and it has only  one existing component.}. In their most general form, the two (1 + 1)-dimensional (2D) gauge-fixed Lagrangian 
densities for the modified Proca theory [without the fermionic 
(anti-)ghost fields] are as follows\footnote{For the 2D theory, we adopt the convention and notations such 
that the {\it background} flat Minkowskian spacetime manifold is endowed with a metric 
$\eta_{\mu\nu}$ with signatures ($+1, -1$) so that $P \cdot Q = \eta_{\mu\nu}\,P^\mu\, Q^\nu
= P_0\,Q_0 - P_1\,Q_1$ is the dot product between two non-null 2D vectors $P_\mu$ and $Q_\mu$. We also choose the antisymmetric
Levi-Civita tensor $\varepsilon_{\mu\nu}$ such that $\varepsilon_{01} = +1 = \varepsilon^{10}$,  
$\varepsilon_{\mu\nu}\, \varepsilon^{\mu\nu} = - 2!, \; \varepsilon_{\mu\nu}\, \varepsilon^{\nu\lambda} = \delta^\lambda_\mu$, etc.}
(see, e.g. [3]). 
\begin{eqnarray*}
{\cal L}_{(b_1)} &=& {\cal B}\,(E - m\, \tilde \phi) - \frac {1}{2}\;{\cal B}^2 + m\, E\, \tilde\phi - \frac{1}{2}\, \partial_\mu \tilde \phi\,
\partial^\mu \tilde \phi +  \frac{1}{2} \;m^2 \, A_\mu\, A^\mu \nonumber\\ &-&
 m\, A_\mu\, \partial^\mu \phi + \frac{1}{2}\, \partial_\mu \phi\,
\partial^\mu \phi 
+ B\,(\partial \cdot A + m\,\phi) + \frac{1}{2}\;B^2,
\end{eqnarray*}
\begin{eqnarray}
{\cal L}_{(b_2)} &=& \bar {\cal B}\,(E + m\, \tilde \phi) - \frac{1}{2}\; \bar{\cal B}^2 - m\, E\, \tilde\phi - \frac{1}{2}\, \partial_\mu \tilde \phi\,
\partial^\mu \tilde \phi + \frac{1}{2}\; m^2 \, A_\mu\, A^\mu  \nonumber\\
&+& m\, A_\mu\, \partial^\mu \phi + \frac{1}{2}\, \partial_\mu \phi\,
\partial^\mu \phi 
+ \bar B\,(\partial \cdot A - m\,\phi) + \frac{1}{2} \; \bar B^2,
\end{eqnarray}   
where ($B, \,\bar B,\, {\cal B},\, \bar{\cal B}$) are  the Nakanishi-Lautrup 
type auxiliary fields and $\tilde \phi$ is a pseudo-scalar field
that has been incorporated in the theory on mathematical as well as physical grounds [2,3].
It will be noted that the (pseudo-)scalar 
fields ($\tilde \phi$)$\phi$ have been added/subtracted in a symmetrical fashion
to the kinetic and gauge-fixing terms, respectively, so that 
we would have appropriate discrete symmetry transformations  in the theory [cf. (89) below].

Let us discuss  the dual-gauge transformations $\delta^{(1,2)}_{dg}$:
\begin{eqnarray}
&& \delta^{(1,2)}_{dg}\, A_\mu = - \varepsilon_{\mu\nu}\, \partial^\nu \Sigma, \qquad
\delta^{(1,2)}_{dg} \,(\partial \cdot A \pm m\, \phi) = 0,\quad \delta^{(1,2)}_{dg}\, \phi = 0,\nonumber\\
&& \delta^{(1,2)}_{dg}\, \tilde \phi = \mp \,m\, \Sigma, \qquad \delta^{(1,2)}_{dg}\,E = \Box\; \Sigma, \qquad
 \delta^{(1,2)}_{dg}\,[B,\, \bar B, {\cal B},\, \bar{\cal B}] = 0,
\end{eqnarray}
where $\Box = \partial^2_0 - \partial^2_1$ is the d'Alembertian operator, $\Sigma(x)$ is the local 
and infinitesimal  dual-gauge parameter  and the superscripts ($1,2$) 
denote the dual-gauge transformations for the Lagrangian densities
${\cal L}_{(b_1)}$ and ${\cal L}_{(b_2)}$, respectively. We note that 
the Lagrangian densities ${\cal L}_{(b_1, b_2)}$ transform, under the above dual-gauge transformations
 $\delta_{dg}^{(1, 2)}$, as follows:
\begin{eqnarray}
\delta_{dg}^{(1)}{\cal L}_{(b_1)} &=& \partial_\mu \;\Bigl[m\, \varepsilon^{\mu\nu}\;(m \,A_\nu\, \Sigma + \phi\, \partial_\nu\;\Sigma) + m\,\tilde\phi\, \partial^\mu\;\Sigma \Bigr] 
+ {\cal B}\;(\Box + m^2)\;\Sigma ,\nonumber\\
\delta_{dg}^{(2)}{\cal L}_{(b_2)} &=& \partial_\mu \;\Bigl[m\; \varepsilon^{\mu\nu}\;(m A_\nu \;\Sigma - \phi \;\partial_\nu\;\Sigma)
- m\tilde\phi\; \partial^\mu\;\Sigma \Bigr] 
+ \bar {\cal B}\;(\Box + m^2)\;\Sigma.
\end{eqnarray}
Thus, it is clear that, to maintain the dual-gauge symmetries in the 2D gauge-fixed 
theory, we have to impose the condition $(\Box + m^2)\,\Sigma (x) = 0$, from outside, on the dual-gauge parameter $\Sigma (x)$.
We note that the operation of co-exterior derivative $\delta = - \,* \,d \,*$
on the connection 1-form ($A^{(1)} = dx^\mu A_\mu$) yields $(\partial \cdot A)$ which is a zero-form. We can
add/subtract a scalar field $\phi$ to it as is the case with 
the gauge-fixing terms $(\partial \cdot A \pm m \phi)$ that have been incorporated in 
${\cal L}_{(b_1, b_2)}$. This scalar field $\phi$ is nothing but the Stueckelberg field.

A few noteworthy points, at this juncture, are as follows. First, we point out 
that the nomenclature of the dual-gauge symmetry 
is appropriate because we have: $\delta_{dg}^{(1, 2)}\,(\partial\cdot A \pm m\phi) = 0$. In other 
words, the total gauge-fixing terms
$(\partial\cdot A \pm m\phi)$, owing their {\it fundamental} origin to the 
dual-exterior derivative, remain invariant. Second, the dual-gauge
parameter has to be restricted by $(\Box + m^2)\, \Sigma = 0$ to maintain the dual-gauge symmetry in the theory.  One can take care
of this restriction by introducing  the (anti-)ghost fields $(\bar C)C$ within the framework of BRST formalism as we shall see in our subsection {\bf 4.3}.
Third, the dual-gauge symmetry transformations exist only in the specific two (1 + 1)-dimensions of spacetime for the Abelian 1-form gauge theory
whereas the local gauge as well as (anti-)BRST symmetries exist in any arbitrary  dimension of spacetime.
Fourth, the {\it perfect} analogue of the gauge symmetry
[cf. (4)] does {\it not} exist for the dual-gauge symmetry (because we have to impose the restriction 
$(\Box + m^2)\, \Sigma = 0$ from outside). Finally, in the forthcoming sections, we shall see that 
one can have {\it perfect} (anti-)dual-BRST [or (anti-)co-BRST] symmetries in the theory
where $\Sigma$ will be replaced by the (anti-)ghost fields $(\bar C)C$ (without any
ad-hoc  restrictions from outside).

We claim that there  would {\it not} be any restrictions on anything (from outside) 
when we shall discuss the full (anti-)BRST and (anti-)co-BRST invariant Lagrangian 
densities of our present theory.

\subsection{Nilpotent (anti-)co-BRST symmetry   transformations: geometrical  superfield
  technique}

As prescribed by the superfield formalism, first of all, we generalize the 2D theory onto the
(2, 2)-dimensional supermanifold and promote the ordinary co-exterior derivative $\delta = - *\,d \,*$ onto
the same supermanifold, as 
\begin{eqnarray}
\delta  = - *\,d \,* \qquad \Longrightarrow \qquad \tilde \delta = - \star \; \tilde d \; \star,
\end{eqnarray} 
where the ($\star$) operator is the Hodge duality operation, defined on the (2, 2)-dimensional supermanifold. The working rule for
the operation of ($\star$) has been worked out in our earlier paper [31] and we exploit these results in the following dual-HC (DHC):
\begin{eqnarray}
\tilde \delta\, \tilde A^{(1)} = \delta\, A^{(1)}, \qquad \quad \delta\,A^{(1)} = (\partial\cdot A),
\end{eqnarray}
where the l.h.s. is ($- \star\, \tilde d \, \star\, \tilde A^{(1)}$) and r.h.s. is obviously equal 
to the Lorentz condition for the gauge-fixing  ($\partial \cdot A$). The definition of
$\tilde d$ and $\tilde A^{(1)}$ are quoted in (6) and the super-expansions   of the superfields are listed in (7).

The explicit expression for the computation of the l.h.s. of  the DHC, in equation (45), 
is as follows\footnote{We have performed explicit computation of ($- \star\, \tilde d \, \star\, {\tilde A}^{(1)}$)
in our Appendix A and derived clearly the equation (46)
which plays an important role in our further discussions.} (see, e.g. [31] for details)
\begin{eqnarray}
\partial \cdot {\cal B} - (\partial_\theta \bar F + \partial_{\bar\theta} F ) - S^{\theta\theta}\, (\partial_\theta F)
- S^{\bar\theta\bar\theta}\, (\partial_{\bar\theta} \bar F),
\end{eqnarray}
where $(S^{\theta\,\theta},\,S^{\bar\theta\,\bar\theta})$ coefficients, in the above, have turned up while taking the
Hodge duality ($\star$) operation on the following  super 4-forms (defined on the (2, 2)-dimensional supermanifold) while the computations of 
$d\star A^{(1)}$ is performed, namely; 
\begin{eqnarray}
\star\, (dx^\mu \wedge dx^\nu \wedge d\theta \wedge d\theta) = \varepsilon^{\mu\nu}\, S^{\theta\,\theta}, \quad
\star\, (dx^\mu \wedge dx^\nu \wedge d \bar\theta \wedge d \bar\theta) = \varepsilon^{\mu\nu}\, S^{\bar\theta\,\bar\theta}.
\end{eqnarray}
It is to be noted that ($\tilde d\, \star\, \tilde A^{(1)}$) is a super 4-form on the (2, 2)-dimensional supermanifold
and when we perform another ($\star$) operation on it,
the differentials of (47) appear.
In the above, $S^{\theta\,\theta}$ and $S^{\bar\theta\,\bar\theta}$ are symmetric in $\theta$ and $\bar\theta$ 
and all the other coefficients of the l.h.s. of (45) have been worked out in our earlier paper [31]. On the comparison
of the l.h.s. and r.h.s. of (45), we obtain:
\begin{eqnarray}
&&\partial \cdot  R^{(1)} = 0, \;\;\qquad \partial \cdot  R^{(2)} = 0, \;\;\qquad \partial \cdot S = 0, \nonumber\\
 &&s = \bar s = 0,\qquad B_1 = B_4 = 0, \qquad  B_2 + B_3 = 0.
\end{eqnarray}
It is clear that, unlike the HC where all the secondary fields of expansions (7) are exactly and uniquely determined,
in the case of DHC, the secondary fields are not uniquely determined and there can be various 
(non-)local choices for the solution of (48) (see, e.g. [32] for details). Thus, we have the complete freedom to make the choices.
Finally, we select\footnote{By exploiting the {\it augmented} version of superfield formalism, we
have derived these {\it exact} expressions in our Appendix B. Thus, results of (49) are
mathematically  precise and exact.} 
the following {\it local} expressions
 for the solution of (48), namely;
\begin{eqnarray}
&& R_\mu^{(1)} = - \varepsilon_{\mu\nu} \, \partial^\nu  C,  
\qquad  R_\mu^{(2)} = - \varepsilon_{\mu\nu} \, \partial^\nu \bar C, \nonumber\\ 
&& S_\mu =  \varepsilon_{\mu\nu} \, \partial^\nu  {\cal B},\qquad  B_3 = - {\cal B}, \qquad B_2 = {\cal B},
\end{eqnarray}
which, unambiguously, satisfy $\partial \cdot  R^{(1)} =  \partial \cdot  R^{(2)} = \partial \cdot S = 0$ and $B_2 + B_3 = 0$.
From now on, we shall focus only
on the Lagrangian density ${\cal L}_{(b_1)}$ of (41) and its generalization  to the 
(anti-) co-BRST invariant Lagrangian density (57)  (see below). 
 However, it is straightforward to make the local choices
for the Lagrangian density ${\cal L}_{(b_2)}$, too. For instance, we can choose 
$R^{(1)}_\mu = - \varepsilon_{\mu\nu}\, \partial^\nu\, C,$ $R^{(2)}_\mu = - \varepsilon_{\mu\nu}\, \partial^\nu\, \bar C,\;
S_\mu = \varepsilon_{\mu\nu}\, \partial^\nu\, \bar{\cal B}$, $B_3 = -\, \bar{\cal B}, \; B_2 = \bar{\cal B}$ 
for the (anti-)co-BRST inavriant version of  ${\cal L}_{(b_2)}$.

Ultimately, we obtain the following expansions for the superfields along the Grassmannian 
($\theta, \bar\theta$)-directions of the (2, 2)-dimensional supermanifold after the application of DHC, namely;
\begin{eqnarray}
{\cal B}^{(dh)}_\mu (x, \theta, \bar\theta) &=& A_\mu + \theta \;(- \varepsilon_{\mu\nu}\;\partial^\nu C)
+  \bar \theta \;(- \varepsilon_{\mu\nu}\;\partial^\nu \bar C) 
+  \theta \bar\theta\;( i\;\varepsilon_{\mu\nu}\;\partial^\nu {\cal B})\nonumber\\
&\equiv & A_\mu + \theta \;(s_{ad} \; A_\mu)
+  \bar \theta \;(s_d \; A_\mu)  +  \theta \bar\theta\;(s_d s_{ad} \; A_\mu),\nonumber\\
F^{(dh)} (x, \theta, \bar\theta) &=& C + \theta \;(0) + \bar\theta\;(-i {\cal B}) 
+ \theta\bar\theta\;(0)\nonumber\\
&\equiv & C+ \theta \;(s_{ad} \;C)+  \bar \theta \;(s_d \; C) +  \theta \bar\theta\;(s_d s_{ad} \; C),\nonumber\\
\bar F^{(dh)} (x, \theta, \bar\theta) &=& \bar C + \theta \;(i {\cal B}) + \bar\theta\;(0) + \theta\bar\theta\;(0)\nonumber\\
&\equiv & \bar C+ \theta \;(s_{ad} \;\bar C)+  \bar \theta \;(s_d \; \bar C) 
+ \theta \bar\theta\;(s_d s_{ad} \; \bar C),
\end{eqnarray}  
where the superscript ($dh$) denotes the expansions of the superfields after the application of  
DHC. A close look at the above expansions demonstrates
that we have already obtained the (anti-)co-BRST symmetry transformations for the gauge field $(A_\mu)$
and corresponding (anti-)ghost fields $(\bar C)C$. Physically, the DHC states that the dual-gauge invariant quantity
[i.e. $\delta^{(1,2)}_{dg}\, (\partial \cdot A = 0)$], which is nothing but the Lorentz condition ($\partial \cdot A$)
for the gauge-fixing, does {\it not} depend on the Grassmannian variables $\theta$ and $\bar \theta$ in any form.

To obtain the (anti-)co-BRST symmetry transformations for the $\tilde\phi$ field, we exploit the strength of augmented superfield
formalism where we demand that all the dual-gauge [or (anti-)co-BRST] invariant quantities should remain independent of the
 Grassmannian variables $\theta$ and $\bar\theta$. In this context, we observe that
 $\delta^{(1)}_{dg}[A_\mu(x) $ $-\,(1/m)\, \varepsilon_{\mu\nu}\partial^\nu \tilde\phi (x)] = 0$ 
under the dual-gauge  transformations (42). Thus, we demand the following dual-GIR on the superfields of the (2, 2)-dimensional 
supermanifold,  namely; 
\begin{eqnarray} 
&&{\cal B}_\mu^{(dh)} \; (x, \theta, \bar\theta) - \, \frac{1}{m}\, \varepsilon_{\mu\nu}\,\partial^\nu\; \tilde\Phi(x, \theta, \bar\theta) 
 = A_\mu (x) - \, \frac{1}{m}\, \varepsilon_{\mu\nu}\partial^\nu\;\tilde\phi(x).
\end{eqnarray} 
We note that the DGIR combines DHC and the dual-gauge invariance {\it together} in a fruitful fashion.
Taking the  help from (50) and using the following expansion for the superfield $\tilde\Phi(x, \theta, \bar\theta)$
along the Grassmannian ($\theta, \bar\theta$)-directions of the (2, 2)-dimensional supermanifold:
\begin{eqnarray} 
\tilde\Phi \; (x, \theta, \bar\theta)  = \tilde \phi(x) + \theta\; f_1 (x) + \bar\theta \; f_2(x) + i\,\theta\,\bar\theta\; b_1(x),
\end{eqnarray} 
we obtain the following results:
\begin{eqnarray} 
 && f_1 (x) = (- \, m\;C), \quad f_2 (x) = (- \,m\;\bar C), \quad
  b_1 (x) = (m\;\cal B).
\end{eqnarray} 
It is obvious, from the above, that $f_1(x)$ and $f_2(x)$ are fermionic in nature and $b_1(x)$ is bosonic.
Plugging in the above values into (52), we deduce the following:
\begin{eqnarray} 
\tilde\Phi^{(dg)}  (x, \theta, \bar\theta)  &=& \tilde {\phi} (x) + \theta\;(- \,m\, C) 
+ \bar\theta \;(- \,m\, \bar C) 
+ \theta\bar\theta\;(i\,m\, \cal B) \nonumber\\
&\equiv & \tilde \phi + \theta\;\bigl(s_{ad}\; \tilde \phi \bigr) + \bar\theta \;\bigl(s_d \;\tilde \phi \bigr) 
+ \theta\bar\theta\; \bigl( s_d\; s_{ad}\; \tilde \phi \bigr),
\end{eqnarray} 
where the superscript ($dg$) denotes the super-expansion after the application of dual-GIR (DGIR) on the superfields
of the (2, 2)-dimensional supermanifold.

A careful observation of (50) and (54) leads to the derivation of the following fermionic (anti-)co-BRST 
symmetry transformations for the whole theory, namely; 
\begin{eqnarray}
&& s_{ad} \, A_\mu = - \varepsilon_{\mu\nu} \, \partial^\nu C, \,\qquad s_{ad}\, C = 0,
\,\qquad s_{ad}\,\bar C = i\, {\cal B}, \quad s_{ad}\, {\cal B} = 0, \nonumber\\ 
&& s_{ad}\,(\partial\cdot A) = 0, \qquad s_{ad}\,\phi = 0, \quad
 s_{ad}\,\tilde \phi  = - \, m \,C, \qquad s_{ad} \,E = \Box\, C, \nonumber\\ 
&& s_d\, A_\mu = - \varepsilon_{\mu\nu} \, \partial^\nu {\bar C},\; \quad s_d\, \bar C = 0, \;\;\quad 
s_d\, C = -i\, {\cal B},\;\; \quad s_d\, {\cal B} = 0, \nonumber\\ 
&& s_d\,(\partial\cdot A) = 0,\quad s_d \, \phi = 0, \qquad s_d \,\tilde \phi  
= - \, m \,\bar C,\qquad s_d \,E = \Box \,\bar C,
\end{eqnarray}
which would be the symmetry transformations for the appropriately modified [cf. (57) below] 
form of the Lagrangian density (21).
A careful observation at the transformations (55) demonstrates that the (anti-)co-BRST 
symmetry transformations are off-shell
nilpotent as we do {\it not} use  any equation of motion in the proof of $s^2_{(a)d} = 0$. Further, these transformations
are absolutely anticommuting in nature because it can be checked that $s_d\,s_{ad} + s_{ad}\,s_d = 0$.
 Finally, we have christened the transformations (55) as (anti-)dual-BRST [or
(anti-)co-BRST] transformations because the total gauge-fixing term ($\partial \cdot A + m\, \phi$),
originating basically from the co-exterior derivative, remains invariant under the nilpotent
($s^2_{(a)d} = 0$) and absolutely anticommuting ($s_d\,s_{ad} + s_{ad}\,s_d = 0$) transformations $s_{(a)d}$.

It is clear that the 2D nilpotent
(anti-)co-BRST symmetry transformations (55) are derived from the super-expansions (50)
and (54) which are present on the (2, 2)-dimensional supermanifold. 
Hence, there should be some connection between the 2D 
nilpotent (anti-)co-BRST symmetries and the superfield formalism on (2, 2)-dimensional superspace. A careful observation,
at the super-expansions in (50) and (54), leads to the following relationships:
\begin{eqnarray}
&&\lim_{\bar \theta = 0}\; \frac{\partial}{\partial \theta}\; \tilde N^{(dh, dg)}\, (x, \theta, \bar\theta) 
=  s_{ad}\; N(x), \qquad \qquad \partial_\theta\; \longleftrightarrow  \; s_{ad}, \nonumber\\
&&\lim_{\theta = 0}\; \frac{\partial}{\partial \bar\theta}\; \tilde N^{(dh, dg)} (x, \theta, \bar\theta)
 =  s_d\; N(x),  \qquad\qquad\;\; \partial_{\bar\theta} \; \longleftrightarrow  \; s_{d},\nonumber\\
&&\frac{\partial}{\partial \bar\theta}\; \frac{\partial}{\partial \theta}\; \tilde N^{(dh, dg)}\, (x, \theta,
 \bar\theta) =  s_d\,s_{ad}\; N(x),  \qquad \partial_{\bar\theta}\,\partial_\theta\; \longleftrightarrow  \; s_d\, s_{ad},
\end{eqnarray}
where ${\tilde N}^{(dh, dg)} \,(x, \theta, \bar\theta)$ is the generic superfield obtained after the application 
of  DHC and DGIR on the (2, 2)-dimensional supermanifold and $N (x)$ is the ordinary 2D field of our present
(anti-)co-BRST invariant theory. It is evident that the transformations (55) would be automatically off-shell
nilpotent and absolutely anticommuting because these are identified with the translational operators 
($\partial_\theta, \partial_{\bar\theta}$), along the Grassmannian directions ($\theta, \bar\theta$) 
of the (2, 2)-dimensional supermanifold, which satisfy 
$\partial_\theta^2 = \partial_{\bar\theta}^2 = 0,\, \partial_\theta\,\partial_{\bar\theta}
+ \partial_{\bar\theta}\,\partial_\theta =0$ due to their inhernt properties. We end
 this subsection with the remark that
$\Phi\,(x, \theta, \bar\theta) = \phi (x)$ because $\phi(x)$ is a dual-gauge invariant 
quantity (i.e. $s_{(a)d}\, \phi = 0$).


\subsection{Lagrangian densities: (anti-)co-BRST invariance}

The (anti-)co-BRST invariant version of the 2D Lagrangian density ${\cal L}_{(b_1)}$ of (41),  is the one that incorporates the FP-ghost terms, namely; 
\begin{eqnarray}
{\cal L}_{\cal B} &=& {\cal B}\,(E - m\, \tilde \phi) - \frac {1}{2}\;{\cal B}^2 + m E \tilde\phi - \frac{1}{2}\, \partial_\mu \tilde \phi\,
\partial^\mu \tilde \phi +  \frac{1}{2} \;m^2  A_\mu A^\mu 
-  m\, A_\mu\, \partial^\mu \phi \nonumber\\&+& \frac{1}{2}\, \partial_\mu \phi\,
\partial^\mu \phi + \frac{1}{2}\;B^2 + B\,(\partial \cdot A + m\,\phi)  - i\,\partial_\mu \,\bar C\, \partial^\mu\, C + i\,m^2\,\bar C\, C.
\end{eqnarray}
We note that the gauge-fixing and Faddeev-Popov ghost terms of the above Lagrangian density
are same as that of the (anti-)BRST invariant Lagrangian density (21). Under the off-shell nilpotent 
and absolutely anticommuting (anti-)co-BRST symmetry transformations $s_{(a)d}$ [cf. (55)], the above
Lagrangian density transforms to the total spacetime derivatives as illustrated below:
\begin{eqnarray}
s_d\, {\cal L}_{\cal B} &=& \partial_\mu \Big[ {\cal B}\, \partial^\mu \bar C +
 m\, \varepsilon^{\mu\nu} \Big( m\, A_\nu\, \bar C
 + \phi\, \partial_\nu \bar C \Big) 
+ m\, \tilde \phi\, \partial^\mu \bar C  \Big], \nonumber\\
s_{ad}\, {\cal L}_{\cal B} &=& \partial_\mu \Big[ {\cal B}\, \partial^\mu C +
m\, \varepsilon^{\mu\nu} \Big( m\, A_\nu\, C
 + \phi\, \partial_\nu C \Big) 
+ m\, \tilde \phi\, \partial^\mu C  \Big].
\end{eqnarray}
Hence, the action integral $S =\int dx\,{\cal L}_{\cal B} $ of our theory  
remains invariant.

We note that the gauge-fixing and  FP-ghost terms of the Lagrangian densities (21) have been derived by exploiting 
the off-shell nilpotent (anti-)BRST symmetry transformations [cf. (20)]. In exactly similar fashion, it is interesting to observe that 
\begin{eqnarray}
 && s_d\,s_{ad}\, \Big[\frac{i}{2}\,A_\mu\,A^\mu + \frac{i}{2}\, \tilde\phi^2 + \frac{1}{2}\,C\,\bar C\Big]\nonumber\\
&&= {\cal B}\, (E - m\,\tilde\phi) - \frac{1}{2}\, {\cal B}^2 
- i\,\partial_\mu \,\bar C\, \partial^\mu\, C + i\,m^2\,\bar C\, C,\nonumber\\
  && s_d\, \Big[(-i)\bigl\{ \varepsilon^{\mu\nu}\, A_\mu\,\partial_\nu\, C + m\, \tilde\phi\, C
+ \frac{1}{2} {\cal B}\, C \bigr \}\Big]\nonumber\\
&& = {\cal B}\, (E - m\,\tilde\phi) - \frac{1}{2}\, {\cal B}^2 
- i\,\partial_\mu \,\bar C\, \partial^\mu\, C  + i\,m^2\,\bar C\, C,\nonumber\\ 
 && s_{ad}\,\Big[(+i)\bigl\{ \varepsilon^{\mu\nu}\, A_\mu\,\partial_\nu\, \bar C + m\, \tilde\phi\, \bar C 
+ \frac{1}{2} {\cal B}\, \bar C  \bigr \} \Big] \nonumber\\ 
&& = {\cal B}\, (E - m\,\tilde\phi) - \frac{1}{2}\, {\cal B}^2 
- i\,\partial_\mu \,\bar C\, \partial^\mu\, C  +\, i\,m^2\,\bar C\, C.
\end{eqnarray}
The above expressions show that there are {\it three} different ways 
(modulo a total spacetime derivative term) to write the kinetic term plus the
FP-ghost terms in the language of the off-shell nilpotent  ($s_{a(d)}^2 = 0$) 
(anti-)co-BRST symmetry transformations  $s_{a(d)}$ which also absolutely
anticommute (i.e. $s_d\,s_{ad} + s_{ad}\,s_d = 0$) with each-other in their operator form.

We can express the above three relations in the language of superfield formalism 
because we observe that the following expressions
\begin{eqnarray}
&&\frac{\partial}{\partial\bar\theta}\,\frac{\partial}{\partial\theta}
\Bigl[ \frac{i}{2} \,{\cal B}_\mu^{(dh)}\, {\cal B}^{\mu (dh)}
 + \frac{i}{2} \,\tilde\Phi^{(dg)}\,\tilde\Phi^{(dg)} + \frac{1}{2}\, F^{(dh)}\,\bar F^{(dh)}\Bigr] \nonumber\\
&&\lim_{\theta = 0}\, \frac{\partial}{\partial \bar\theta}\,\Bigl[ (-i)\; \{\varepsilon^{\mu\nu} {\cal B}_\mu^{(dh)}\,
\partial_\nu\, F^{(dh)} + m\, \tilde\Phi^{(dg)} F^{(dh)}  
 + \frac{1}{2}\,{\cal B}(x)\, F^{(dh)} \}\Bigr] \nonumber\\
&&\lim_{\bar\theta = 0}\, \frac{\partial}{\partial \theta}\,\Bigl[ (+i)\; \{\varepsilon^{\mu\nu} {\cal B}_\mu^{(dh)}\,
\partial_\nu\, {\bar F}^{(dh)} + m\, \tilde\Phi^{(dg)} {\bar F}^{(dh)}  
+ \frac{1}{2}\,{\cal B}(x)\,{\bar F}^{(dh)} \}\Bigr],
\end{eqnarray}
also lead to the derivation of the sum of a part of kinetic term and FP-ghost terms. 
Ultimately,  this exercise implies that the sum of kinetic  and FP-ghost terms
 \begin{eqnarray}
{\cal B}\, (E - m\,\tilde\phi) - \frac{1}{2}\, {\cal B}^2 
- i\,\partial_\mu \,\bar C\, \partial^\mu\, C + i\,m^2\,\bar C\, C,
\end{eqnarray}
is {\it always} (anti-)dual-BRST invariant quantity (modulo a total spacetime derivative) 
because this is trivially  true when we take into account the nilpotency and absolute 
anticommutativity of the (anti-)co-BRST symmetry transformations. In other words,
we conclude that $s_{(a)d}\,[{\cal B} (E - m\, \tilde \phi) - (1/2) {\cal B}^2
- i\,\partial_\mu \,\bar C\, \partial^\mu\, C + i\,m^2\,\bar C\, C]$ $= 0$ modulo a total spacetime derivative.
Thus, a part of Lagrangian density (57) is invariant under $s_{(a)d}$.

We have already seen that a part of the kinetic term and the total of 
FP-ghost terms can be expressed in terms of the superfields 
obtained after the application of DHC and DGIR [cf. (60)]. Furthermore, the kinetic term
 $\frac{1}{2}\,\partial_\mu\phi\,\partial^\mu\phi$ for the $\phi$ field and the total gauge 
fixing term [$B\, (\partial\cdot A+ m\, \phi) + 1/2(B^2)$] would remain intact within the framework 
of superfield formalism as they are the dual-gauge [or (anti-)co-BRST] invariant quantities.
We note that $B\,(\partial\cdot A + m\,\phi) \rightarrow B\, (\partial_\mu
{\cal B}^{\mu(dh)} + m\phi)$ in the superfield formalism and it is trivial to check that 
$B\,(\partial\cdot{\cal B}^{(dh)}) = B\, (\partial\cdot A)$ [cf. (50)] so that
$B\,(\partial \cdot A + m\, \phi) \to B\,(\partial \cdot A + m\, \phi)$ 
without any change whatsoever when we generalize it onto the (2, 2)-dimensional supermanifold.
The rest of the terms can be generalized onto the (2, 2)-dimensional supermanifold as
\begin{eqnarray}
- m\,\varepsilon^{\mu\nu}\, \partial_\mu\, {\cal B}_\nu^{(dh)}\, \tilde\Phi^{(dg)} 
- \frac{1}{2}\, \partial_\mu \tilde\Phi^{(dg)}\, \partial^\mu \tilde\Phi^{(dg)} \nonumber\\ + \frac{1}{2}\, m^2\,{\cal B}_\mu^{(dh)}\,
{\cal B}^{\mu(dh)}\, - m\, {\cal B}_\mu^{(dh)}\,\partial^\mu\phi(x),
\end{eqnarray} 
where the symbols have already been explained earlier and they are nothing but 
the super-expansions after the application of the DHC and DGIR
[cf. (50), (54)].

It is interesting to note that the 
{\it last} term of (62) is {\it always} an (anti-)co-BRST invariant quantity because
we observe the following:
\begin{eqnarray}
- m {\cal B}_\mu^{(dh)}\, \partial^\mu\phi 
 = - m[ A_\mu + \theta (- \varepsilon_{\mu\nu} \partial^\nu C)
+  \bar \theta (- \varepsilon_{\mu\nu}\partial^\nu \bar C) 
+  \theta \bar\theta( i\varepsilon_{\mu\nu} \partial^\nu {\cal B})]\,\partial^\mu\phi,
\end{eqnarray} 
where we have taken the expansions of ${\cal B}_\mu^{(dh)}(x, \theta, \bar\theta)$ from (50). 
Taking the help of the mappings (56), we note the following
\begin{eqnarray}
\lim_{\theta  = 0}\, \frac{\partial}{\partial\bar\theta}\,\Bigl[- m\, {\cal B}_\mu^{(dh)}\, \partial^\mu\phi\Bigr] 
 &=& \partial_\mu \Bigl[ m \; \varepsilon^{\mu\nu}\;\phi\,\partial_\nu \bar C\Bigr] 
\equiv s_d\,\Bigl[-m\, A_\mu\,\partial^\mu\phi\Bigr],\nonumber\\
\lim_{\bar\theta = 0}\, \frac{\partial}{\partial\theta}\,\Bigl[- m\, {\cal B}_\mu^{(dh)}\, \partial^\mu\phi\Bigr] 
 &=& \partial_\mu \Bigl[ m \; \varepsilon^{\mu\nu}\;\phi\,\partial_\nu  C\Bigr] 
\equiv s_{ad}\,\Bigl[-m\, A_\mu\,\partial^\mu\phi\Bigr], \nonumber\\
 \frac{\partial}{\partial\bar\theta}\, \frac{\partial}{\partial\theta}\Bigl[- m\, {\cal B}_\mu^{(dh)} \partial^\mu\phi\Bigr] 
 &=& \partial_\mu \Bigl[- i m  \varepsilon^{\mu\nu}\phi\partial_\nu {\cal B}\Bigr] 
\equiv s_d s_{ad} \Bigl[-m A_\mu\partial^\mu\phi\Bigr],
\end{eqnarray} 
which demonstrates that $ s_{(a)d}\,[-m\, A_\mu\,\partial^\mu\phi]$ is always a total spacetime derivative.
The rest of the terms in (62) are also (anti-)co-BRST invariant quantity because we check that 
\begin{eqnarray*}
&&\lim_{\theta = 0}\, \frac{\partial}{\partial\bar\theta}\,\Bigl[- m\,\varepsilon^{\mu\nu}\,
\partial_\mu {\cal B}_\nu^{(dh)}\, \tilde\Phi^{(dg)} - \frac{1}{2}\, \partial_\mu \tilde\Phi^{(dg)}\, \partial^\mu \tilde\Phi^{(dg)} \nonumber\\
&&+ \frac{1}{2}\, m^2\,{\cal B}_\mu^{(dh)}\,
{\cal B}^{\mu(dh)}\Bigr] = \partial_\mu \Bigl[ m \,\tilde\phi\,\partial^\mu \bar C +  m^2  \varepsilon^{\mu\nu}\,A_\nu \,\bar C \Bigr] \nonumber\\  
&&\equiv  s_d \Bigl[ m\, E\, \tilde\phi - \frac{1}{2}\, \partial_\mu \tilde \phi\,
\partial^\mu \tilde \phi + \frac{1}{2} \;m^2 A_\mu A^\mu \Bigr], \nonumber\\ 
&&\lim_{\bar\theta = 0}\, \frac{\partial}{\partial\theta}\,\Bigl[- m\,\varepsilon^{\mu\nu}\,
\partial_\mu {\cal B}_\nu^{(dh)}\, \tilde\Phi^{(dg)} - \frac{1}{2}\, \partial_\mu\tilde\Phi^{(dg)}\, 
\partial^\mu \tilde\Phi^{(dg)} \nonumber\\  
&& + \frac{1}{2}\, m^2\,{\cal B}_\mu^{(dh)}\,
{\cal B}^{\mu(dh)}\Bigr] = \partial_\mu \Bigl[ m \,\tilde\phi\,\partial^\mu   C +  m^2  \varepsilon^{\mu\nu}\, A_\nu \, C\Bigr]  \nonumber\\  
&&\equiv  s_{ad} \Bigl[ m\, E\, \tilde\phi - \frac{1}{2}\, \partial_\mu \tilde \phi\,
\partial^\mu \tilde \phi + \frac{1}{2} \;m^2  A_\mu A^\mu \Bigr], 
\end{eqnarray*}
\begin{eqnarray}
&& \frac{\partial}{\partial\bar\theta}\frac{\partial}{\partial\theta}\,\Bigl[- m\,\varepsilon^{\mu\nu}\,
\partial_\mu {\cal B}_\nu^{(dh)}\, \tilde\Phi^{(dg)} - \frac{1}{2}\, \partial_\mu \tilde\Phi^{(dg)}\, 
\partial^\mu \tilde\Phi^{(dg)} \nonumber\\ 
&&+ \frac{1}{2}\, m^2\,{\cal B}_\mu^{(dh)}\,{\cal B}^{\mu(dh)}\Bigr] = \partial_\mu \Bigl[ m^2\,  (C\,\partial^\mu \, \bar C - \bar C\,\partial^\mu \,C)\nonumber\\  
&& - i\,m\,(\tilde \phi \, \partial^\mu {\cal B} + m\, \varepsilon^{\mu\nu}\; A_\nu \,{\cal B})\Bigr]\nonumber\\
 &&\equiv  s_d\, s_{ad}\,\Bigl[ m\, E\, \tilde\phi - \frac{1}{2}\, \partial_\mu \tilde \phi\,
\partial^\mu \tilde \phi + \frac{1}{2} \;m^2 \, A_\mu\, A^\mu \Bigr].
\end{eqnarray} 
Ultimately, we conclude that $s_{(a)d}\,[ m\, E\, \tilde\phi - \frac{1}{2}\, \partial_\mu \tilde \phi\,
\partial^\mu \tilde \phi + \frac{1}{2} \;m^2 \, A_\mu\, A^\mu ]$ is 
{\it always} a total spacetime derivative.
As a consequence, this specific part of the Lagrangian density (i.e. $m\, E\, \tilde\phi - \frac{1}{2}\, \partial_\mu \tilde \phi\,
\partial^\mu \tilde \phi + \frac{1}{2} \;m^2 \, A_\mu\, A^\mu $) is an (anti-)co-BRST invariant quantity.

Ultimately, we have the total expression for the 2D Lagrangian density  (57) in the superfield formalism, 
on the (2, 2)-dimensional supermanifold, as 
\begin{eqnarray}
{\cal L}_{\cal B} \longrightarrow \tilde {\cal L}_{\cal B} &=& 
\frac{\partial}{\partial\bar\theta}\;\frac{\partial}{\partial\theta} \;\left[ \frac{i}{2} 
\,{\cal B}_\mu^{(dh)}\; {\cal B}^{\mu (dh)}
 + \frac{i}{2} \,\tilde\Phi^{(dg)}\;\tilde\Phi^{(dg)} 
+ \frac{1}{2}\, F^{(dh)}\,\bar F^{(dh)} \right]  \nonumber\\  
&+& \frac{1}{2}\,\partial_\mu \phi\, \partial^\mu\phi  + \frac{1}{2}\, B^2(x)  
 -  m\,\varepsilon^{\mu\nu}\,\partial_\mu {\cal B}_\nu^{(dh)}\, \tilde\Phi^{(dg)} - \frac{1}{2}\,
 \partial_\mu \tilde\Phi^{(dg)}\, \partial^\mu\tilde\Phi^{(dg)} \nonumber\\ &+& \frac{1}{2}\, m^2\,{\cal B}_\mu^{(dh)}\,{\cal B}^{\mu(dh)}
 - m\, {\cal B}_\mu^{(dh)}\, \partial^\mu\,\phi + B(x) \Big(\partial_\mu {\cal B}^{\mu(dh)} + m\, \phi(x) \Big),
 \end{eqnarray}
where all the symbols have  been explained earlier. The (anti-)co-BRST invariance of the Lagrangian density,
 within the framework of superfield formalism, is
\begin{eqnarray}
\lim_{\theta = 0}\, \frac {\partial}{\partial \bar\theta}\; \tilde{\cal L}_{\cal B} 
 &=& \partial_\mu \Big[ {\cal B} \, \partial^\mu\, \bar C +
m\, \varepsilon^{\mu\nu} \phi \,\partial_\nu  \bar C + m\, \tilde\phi\, \partial^\mu \bar C 
+ m^2\, \varepsilon^{\mu\nu} A_\nu  \bar C  \Big] \equiv s_d\, {\cal L}_{\cal B},\nonumber\\
\lim_{\bar\theta = 0}\, \frac {\partial}{\partial \theta}\; \tilde{\cal L}_{\cal B} 
&=& \partial_\mu \Big[ {\cal B} \, \partial^\mu\, C +
m\, \varepsilon^{\mu\nu} \phi \,\partial_\nu  C + m\, \tilde\phi\, \partial^\mu  C 
+ m^2\, \varepsilon^{\mu\nu} A_\nu   C  \Big] \equiv s_{ad}\, {\cal L}_{\cal B},\nonumber\\ 
\frac {\partial}{\partial \bar\theta}\, \frac {\partial}{\partial \theta}\; \tilde{\cal L}_{\cal B} 
&=& - \,\partial_\mu \Bigl[ i\, {\cal B} \, \partial^\mu {\cal B}
+ m^2 \,\partial^\mu \,(\bar C\, C)\nonumber\\ 
&+& i\,m\,\bigl \{\tilde \phi \, \partial^\mu {\cal B} + \varepsilon^{\mu\nu}\; (m \,A_\nu \,{\cal B}
+ \phi \, \partial_\nu \, {\cal B} ) \bigr \} \Bigr]  
 \equiv    s_d\, s_{ad}\,{\cal L}_{\cal B}.
\end{eqnarray}
Due to the above observations, it is clear that the action integral would remain invariant under the nilpotent
(anti-)co-BRST symmetry transformations.

Finally, we would like to state that we have accomplished 
our goal of capturing the (anti-)co-BRST invariance of 
the action integral within the framework of superfield formalism 
where we have used the superfields that have been obtained after the application of DHC and DGIR. 
We further note that the expressions in (58) and (67) match very nicely. The appearance of the terms 
like ${\cal B}\, \partial^\mu \,\bar C, {\cal B}\, \partial^\mu \, C$,  $i\, {\cal B}\, \partial^\mu \, {\cal B}$
in the parenthesis of above equation is due to the same kind of arguments as  offered at the end of equation (29) in 
the context of (anti-)BRST symmetries and corresponding  invariance of the action integral under these 
 continuous and nilpotent symmetry
transformations.

\subsection{Nilpotency and anticommutativity of the conserved (anti-) co-BRST charges:
superfield formulation}

Exploiting the standard technique of the Noether theorem and using the appropriate 
equations of motion, we obtain the following 
expressions for the conserved (${\dot Q}_{(a)d} = 0$) and 
off-shell nilpotent (${\dot Q}^2_{(a)d} = 0$) (anti-)co-BRST [or (anti-)dual BRST] charges:
\begin{eqnarray}
&& Q_{ad} = \int dx\, \bigl[{\cal B}\, \dot  C - \dot {\cal B}\, C \bigr] \equiv \int dx\, J^0_{(ad)}, \quad
 Q_d = \int dx\, \bigl[ {\cal B}\, \dot{\bar C} - \dot {\cal B}\, \bar C \bigr]  \equiv \int dx\, J^0_{(d)} ,
\end{eqnarray}  
which have been derived from the Lagrangian density (57) that has led to the following conserved 
(i.e. $\partial_\mu J^\mu_{(a)d} = 0$) Noether currents
\begin{eqnarray}
J^\mu_{ad} &=& {\cal B}\, \partial^\mu  C -  \varepsilon^{\mu\nu} B\, \partial_\nu  C + m\,  C\, \partial^\mu \tilde \phi - m\, \varepsilon^{\mu\nu} \phi\, \partial_\nu C,\nonumber\\
J^\mu_d &=& {\cal B}\, \partial^\mu \bar C -  \varepsilon^{\mu\nu} B\, \partial_\nu \bar C + m\, \bar C\, \partial^\mu \tilde \phi - m\, \varepsilon^{\mu\nu} \phi\, \partial_\nu \bar C.
\end{eqnarray}
The conservation law (i.e. $\partial_\mu J^\mu_{(a)d} = 0$) can be proven by exploiting the following equations of motion
emerging from the Lagrangian density (57), namely;
\begin{eqnarray}
&&B = - (\partial \cdot A + m \,\phi), \qquad \qquad
\Box\, \tilde\phi - m\,({\cal B}  - E) = 0, \nonumber\\ 
&& (\Box + m^2)\, \bar C =0,\;\; \quad  {\cal B} = E - m\,\tilde\phi, \;\;\quad
 \Box\, \phi - m\,(\partial \cdot A + B) = 0, \nonumber\\
&& (\Box + m^2)\, C = 0, \quad \epsilon^{\mu\nu}\,\partial_\mu ({\cal B} + m\, \tilde \phi) - \partial^\nu B 
+ m^2 A^\nu - m\, \partial^\nu \phi = 0.
\end{eqnarray}
It is straightforward to check that the (anti-)co-BRST charges can be expressed in terms of 
the (anti-)co-BRST symmetry transformations as:
\begin{eqnarray}
 Q_{ad} = \int dx\; \Big[ s_{ad}\, s_d \,\Big( i\,  {A}_1 C \Big)\Big],\qquad
 Q_d = \int dx \;\Big [s_d\,s_{ad}\,\Big( -i\,  {A}_1 \bar C \Big) \Big].
\end{eqnarray}
Exploiting the mapping (56), it can be seen that the above expressions could be recast in the language 
of the superfields, obtained after the application of DHC and DGIR, as
\begin{eqnarray}
Q_{ad} &=& \frac{\partial}{\partial\theta}\frac{\partial}{\partial\bar\theta} \int dx\, \Big[ i\, {\cal B}^{(dh)}_1\, F^{(dh)} \Big] 
\equiv \int dx \int d\theta \int d\bar\theta\, \Big[i\, {\cal B}^{(dh)}_1\, F^{(dh)} \Big], \nonumber\\
 Q_d &=& \frac{\partial}{\partial\bar\theta}\,\frac{\partial}{\partial\theta} \int dx\,
 \Big[ - i\, {\cal B}^{(dh)}_1\, \bar F^{(dh)}\, \Big] 
\equiv  \int dx \int d\theta \int d\bar\theta\, \Big[- i\, {\cal B}^{(dh)}_1\, \bar F^{(dh)} \Big].
\end{eqnarray}
From the above expressions, too, one can prove the off-shell nilpotency ($Q^2_{(a)d} = 0$) 
of the charges $Q_{(a)d}$ by observing that the following are true, namely;
\begin{eqnarray}
&& \lim_{\bar\theta = 0}\,\frac{\partial}{\partial \theta}\; Q_{ad} = 0 
\quad\Longleftrightarrow \quad s_{ad}\, Q_{ad} = 0 \equiv i \, \{Q_{ad}, \, Q_{ad}\}, \nonumber\\
&& \lim_{\theta = 0}\,\frac{\partial}{\partial \bar\theta}\; Q_d = 0 \quad\; \Longleftrightarrow \qquad s_{d}\, Q_{d} = 0 \equiv i\, \{Q_d, \, Q_d\}.
\end{eqnarray}
The above observation of the nilpotency of $Q_{(a)d}$ is intimately connected with the nilpotency
$\partial^2_\theta = \partial^2_{\bar\theta} = 0$ of translational generators ($\partial_\theta, 
\partial_{\bar\theta}$) along the Grassmannian directions of the supermanifold.

The nilpotency of $Q_{(a)d}$ can {\it also} be proven by the following expressions of $Q_{(a)d}$ in terms of 
the (anti-)co-BRST symmetry transformations $s_{(a)d}$, namely;
\begin{eqnarray}
&&Q_{ad} = \int dx\, s_{ad}\,\Big[{\cal B}(x)\, A_1(x) + i\, \dot{\bar C}\,  C \Big], \nonumber\\ 
&&Q_d = \int dx\, s_d\,\Big[{\cal B}(x)\, A_1(x) + i\, \bar C\, \dot C \Big].
\end{eqnarray}
Thus, it is clear that the following will be true, namely;
\begin{eqnarray}
&& s_d\, Q_d = i\, \{Q_d, \; Q_d \} = \int dx\, s^2_d\, \Big[ {\cal B}(x)\, A_1(x) 
+ i\, \dot{\bar C}\, C\Big] = 0, 
\qquad\quad (s_d^2 = 0), \nonumber\\
&& s_{ad}\, Q_{ad} = i\, \{Q_{ad}, \, Q_{ad} \} = \int dx\, s^2_{ad}\,
 \Big[ {\cal B}(x)\, A_1(x) + i\, \bar C\, \dot C\Big] = 0,\;\; (s_{ad}^2 = 0),
\end{eqnarray}
due to the nilpotency of $s_{(a)d}$ (i.e. $s^2_{(a)d} = 0 \Leftrightarrow Q^2_{(a)d} = 0$). In the language of superfield
formalism, the expressions (74) can be written as
\begin{eqnarray}
Q_d &=& \lim_{\theta = 0}\frac{\partial}{\partial\bar\theta}\; \int dx\,
 \Big[{\cal B}(x)\, {\cal B}^{(dh)}_1 (x, \theta, \bar\theta) 
+ i\, \bar F^{(dh)} (x,\theta,\bar\theta)\, {\dot F}^{(dh)} (x, \theta, \bar\theta) \Big], \nonumber\\
Q_{ad} &=& \lim _{\bar\theta = 0}\frac{\partial}{\partial \theta} 
\int dx \Big[{\cal B}(x) {\cal B}^{(dh)}_1(x, \theta, \bar\theta) 
+ i \dot{\bar F}^{(dh)} (x, \theta, \bar\theta) F^{(dh)} (x, \theta, \bar\theta) \Big],
\end{eqnarray}
which demonstrate trivially the following
\begin{eqnarray}
&& \lim_{\theta = 0}\,\frac{\partial}{\partial \bar\theta}\;  Q_d = 0\; \Longleftrightarrow \; s_d\, Q_d = 0, \nonumber\\
&& \lim_{\bar\theta = 0}\,\frac{\partial}{\partial \theta}\;  Q_{ad} = 0 \;\Longleftrightarrow \; s_{ad}\, Q_{ad} = 0,
\end{eqnarray}
where the nilpotency of $\partial_\theta$ and $\partial_{\bar\theta}$ (i.e. $\partial_\theta^2 
= \partial_{\bar\theta}^2  = 0 $)
plays a decisive role.

To prove the absolute anticommutativity of $Q_{(a)d}$, we note the following interesting expressions
for the conserved (anti-)co-BRST charges:
\begin{eqnarray}
&& Q_d = \int dx\,\Big[ s_{ad} \Big (- i\, \bar C\, \dot{\bar C} \Big ) \Big], \qquad
 Q_{ad} = \int dx\,\Big[ s_d \Big ( i\,  C\, \dot C \Big ) \Big].
\end{eqnarray}
The above expressions automatically imply the following beautiful relationships: 
\begin{eqnarray}
s_{ad}\, Q_{d} &=& i\, \{ Q_{d}, Q_{ad}\}  
= \int d x \;\Big[ s^2_{ad}\; \Big(-i\, \bar C\, \dot{\bar C}\Big) \Big] = 0, 
\quad (s_{ad}^2 = 0), \nonumber\\
s_d\, Q_{ad} &=& i\, \{ Q_{ad}, Q_{d}\} 
= \int d x \;\Big[ s^2_{d}\; \Big(i\,  C\, \dot{ C}\Big)\Big] = 0,
\qquad (s_d^2 = 0). 
\end{eqnarray}
Thus, we point out a very interesting observation that the absolute anticommutativity property of the 
(anti-)co-BRST charges is deeply and clearly connected with the nilpotency of the (anti-)co-BRST symmetry transformations (i.e. $s_{(a)d}^2 = 0$). 
These expressions (78) could be also written in terms of superfields, translational generators 
($\partial_\theta,\, \partial_{\bar\theta}$) and differentials ($d\theta,\, d\bar\theta$)
defined on the (2, 2)-dimensional supermanifold, as  
\begin{eqnarray}
Q_{d} &=&  \lim_{\bar\theta = 0}\frac{\partial}{\partial\theta}\; \int d x\; \Big[-i\, \bar F^{(dh)}\, 
\dot{\bar F}^{(dh)} \Big]
\equiv  \int d x\;\int d\theta\, \Big[-i\, \bar F^{(dh)}\, \dot{\bar F}^{(dh)} \Big],\nonumber\\
Q_{ad} &=& \lim_{\theta = 0}\frac{\partial}{\partial\bar\theta}\; \int d x\; \Big[i\,  F^{(dh)}\, \dot{ F}^{(dh)} \Big]
\equiv  \int d x\;\int d\bar\theta\; \Big[i\,  F^{(dh)}\, \dot{ F}^{(dh)} \Big]. 
\end{eqnarray}
The above expressions capture the anticommutativity property
of the (anti-)co-BRST charges in the language of superfield formalism, as
\begin{eqnarray}
\lim_{\bar\theta = 0}\frac{\partial}{\partial\theta}\;  Q_{d} = 0,\qquad\qquad 
\lim_{\theta = 0}\frac{\partial}{\partial\bar\theta}\;  Q_{ad} = 0, 
\end{eqnarray}
where the properties $\partial_\theta^2 = 0,\, \partial_{\bar\theta}^2 = 0$ play important roles
when  we use the expressions for $   Q_{(a)d}$ from (80). The anticommutativity property is hidden
in (81) in view of the mapping (56) which imply that (81) can be written as: $s_d Q_{ad} = i\, \{Q_{ad}, Q_d \} = 0$
and $s_{ad} Q_{d} = i\, \{Q_{d}, Q_{ad} \} = 0$ primarily due to $\partial_\theta^2 = 0,\, \partial_{\bar\theta}^2 = 0$.

We end this subsection with the remark that the nilpotency and absolute anticommutativity properties of the (anti-)co-BRST symmetry transformations (and their
corresponding conserved charges) are related with the properties $\partial_\theta^2 = 0,\, \partial_{\bar\theta}^2 = 0$
and $\partial_\theta \, \partial_{\bar\theta}  + \partial_{\bar\theta} \, \partial_{\theta}= 0$. These relations are, in turn, inter-connected with each-other because the limiting case of the latter (i.e. $\partial_\theta \, \partial_{\bar\theta}  + \partial_{\bar\theta} \, \partial_{\theta}= 0$) leads to the derivation of the former ($\partial_\theta^2 = 0,\, \partial_{\bar\theta}^2 = 0$) when we set $\partial_\theta =  \partial_{\bar\theta}$
in the latter relationship of anticommutativity between $\partial_{\theta}$ and $\partial_{\bar\theta}$.

\section{On a unique bosonic symmetry, the  ghost-scale symmetry and the discrete
 symmetries}

From the four nilpotent ($s^2_{(a)b} = s^2_{(a)d} = 0$) symmetries of the theory, we can 
construct a unique bosonic symmetry $s_\omega = \{s_b, s_d\} \equiv - \{s_{ab}, s_{ad}\}$, under which, 
the relevant fields of our present theory [described by the Lagrangian density (57)] transform as
\begin{eqnarray}
&& s_\omega \, A_\mu =  \varepsilon_{\mu\nu}\, \partial^\nu  B +  \partial_\mu  {\cal B},
\qquad \;\;  s_\omega\, \tilde\phi =  m\, B,\;\;\qquad s_\omega \,\phi =  m\, {\cal B},  \nonumber\\
&&  s_\omega\,(\partial\cdot A) =  \Box \,{\cal B}, \qquad
 s_\omega\, E = -\, \Box\,B, \qquad s_\omega \,(B,\, {\cal B}, \, C, \bar C) = 0,
\end{eqnarray} 
modulo an overall factor of ($-\,i$). We note that $\{s_d, s_{ad}\}$ $ = 0,\; \{s_d, s_{ab}\} = 0,
\;\{s_b, s_{ad}\} = 0,\; \{s_b, s_{ab}\} = 0$. We point out that the fundamental symmetries of the theory
are $s_{(a)b}$ and $s_{(a)d}$ which have been derived from the augmented superfield formalism. The
bosonic symmetry transformation $s_\omega$ is derived from the above {\it basic} off-shell nilpotent 
($s_{(a)b}^2 = s_{(a)d}^2 = 0 $) symmetries $s_{(a)b}$ and $s_{(a)d}$. One of the decisive features of the above bosonic symmetry is the observation
that the ghost part of the Lagrangian density remains invariant.

Under the above transformations (82), the Lagrangian density (57) transforms as
\begin{eqnarray}
s_\omega\,{\cal L}_B = \partial_\mu \Big[B\, \partial^\mu {\cal B} - {\cal B}\, \partial^\mu B -  m\, \tilde \phi\, \partial^\mu B  
 - m\, \varepsilon^{\mu\nu} (\phi\, \partial_\nu B + m\, A_\nu\,B)  \Big].
 \end{eqnarray}
As a consequence, the action integral $S = \int dx\, {\cal L}_{\cal B}$ remains invariant. The above symmetry
transformation, according to Noether's theorem, leads to the derivation of the following conserved charge
(as the analogue of the Laplacian operator):
\begin{eqnarray}
Q_\omega = \int \, dx\, J^0_\omega = \int dx\, \Big[ B \, \dot {\cal B}  - \dot B\, {\cal B}\,\Big], 
\end{eqnarray}
which emerges from the Noether conserved ($ \partial_\mu\, J^\mu_\omega = 0$) current
\begin{eqnarray}
J^\mu_\omega &=& \varepsilon^{\mu\nu}\,\bigl(B\, \partial_\nu \,B - {\cal B}\, \partial_\nu\,{\cal B}-
 m\, \tilde\phi\, \partial_\nu \,{\cal B} + m\, \phi\, \partial_\nu  \, B 
\nonumber\\
&+& m^2\,  A_\nu \, B\bigr) +  m \,{\cal B}\, \partial^\mu  \,\phi 
-  m \, B\, \partial^\mu  \,\tilde\phi - m^2\, A^\mu\,{\cal B}. 
\end{eqnarray}
The conserved charge (84) is the generator of the continuous and infinitesimal 
bosonic symmetry transformations (82) which can be checked by using the standard 
formula between the continuous  symmetry and its generator.

Our theory, described  by the Lagrangian density (57), 
is endowed with the following ghost-scale symmetry transformations
[with a {\it global} (i.e. spacetime independent) scale parameter $\Omega$], namely; 
\begin{eqnarray}
&& C\longrightarrow e^{(+1)\cdot\Omega}\, C, \qquad \bar C\longrightarrow e^{(-1)\cdot\Omega}\, \bar C,\nonumber\\ 
&&\Psi\longrightarrow e^{(0)\cdot\Omega}\,\Psi,  \qquad (\Psi  = A_\mu,\,\phi,\, \tilde\phi,\, B,  \, {\cal B}),
\end{eqnarray}
where the numerals  in the exponentials denote the ghost numbers of the fields. The infinitesimal
version of the above scale transformations ($s_g$), are 
\begin{eqnarray}
&& s_g\, C = + C, \qquad s_g\, \bar C = - \bar C, \qquad s_g\, \Psi = 0, \qquad
 (\Psi = A_\mu, \,\phi, \,\tilde\phi,\, B,\, {\cal B}), 
\end{eqnarray}
modulo a factor of $\Omega$ that can be set equal to {\it one}
for the sake of brevity. The above transformations are generated by the following ghost charge $Q_g$ [2,3]:
\begin{eqnarray}
Q_g = i\,\int\, dx\Big[ \bar C\, \dot C - \dot {\bar C}\, C \Big] \equiv \int\; dx\; J^{0}_g.
\end{eqnarray}
This charge has been derived from the conserved current 
$J^\mu_g = i(\bar C\, \partial^\mu C - \partial^\mu {\bar C} \,C)$. The conservation law $\partial_\mu J^\mu_g = 0$
can be proven by using the  Euler-Lagrange 
equations of motion $(\Box + m^2)\, \bar C = 0$ and $(\Box + m^2)\,C = 0$ which emerge from (57).

In addition to the above continuous symmetries, we have a set of suitable discrete symmetries in the theory.
These symmetries are as follows:
\begin{eqnarray}
&& A_\mu \;\rightarrow \; \pm i\, \varepsilon_{\mu\nu}\, 
A^\nu, \quad\qquad E \;\rightarrow \; \mp\, i (\partial \cdot A), 
\qquad (\partial \cdot A) \;\rightarrow \; \mp i\, E, \nonumber\\ &&
 B \;\rightarrow \;\pm \,i\, {\cal B}, \qquad
{\cal B} \;\rightarrow \; \pm\, i\, B, \qquad
 C \;\rightarrow \;\mp\, i\, \bar C, \qquad \bar C \;\rightarrow \; \mp\,i\, C. 
\end{eqnarray}
It is straightforward to check that the Lagrangian density (57) remains 
invariant under the above discrete symmetry transformations.
Further, it can be readily verified that the following is true, namely;
\begin{eqnarray}
&& * \; Q_b = +\; Q_d, \qquad  * \; Q_d = +\; Q_b, \,\qquad  * \; Q_{\omega} = +\; Q_{\omega},  \nonumber\\
&& * \; Q_{ab} = +\; Q_{ad},\; \quad  * \; Q_{ad} = +\; Q_{ab}, \;\quad * \; Q_g = -\; Q_g, \nonumber\\
&& * \; \;(*\;\;Q_r) = +\; \;Q_r, \qquad  (r = b,\, ab, \, d, \,ad, \, \omega),
\end{eqnarray}
where the operator ($*$) is nothing but the operation of the above discrete symmetry transformations
on the conserved charges of the theory. We note that two successive operations of the discrete symmetry
transformations leave the conserved charges intact. On the other hand, a single operation of the discrete
symmetry transformations interchanges each of the pairs ($Q_b, Q_d$) and ($Q_{ab}, Q_{ad}$) such that:
($Q_b \leftrightarrow Q_d,\, Q_{ab} \leftrightarrow Q_{ad}  $) and 
the ghost charge transforms as: $Q_g \rightarrow\,-\, Q_g$.

\section{Algebraic structures and  cohomological aspects}

It can be checked that the six conserved (i.e. $\dot {Q}_r = 0$) 
charges (i.e. $Q_r, \,r = b, ab, d, ad, \omega, g$)  of the theory
obey the following extended BRST algebra:
\begin{eqnarray}
&& Q_{(a)b}^2 = 0,\;\; \qquad Q_{(a)d}^2 = 0,\;\; \qquad \{Q_b,\; Q_{ab}\} 
= 0,\;\;\qquad \{Q_b,\; Q_{ad}\} = 0, \nonumber\\ && 
\{ Q_d,\; Q_{ad} \} = 0,\qquad\qquad \; \{ Q_d,\; Q_{ab}\} 
= 0,\;\qquad \qquad i\;[Q_{g}, \; Q_d] = -\; Q_d,\nonumber\\ 
&& i\;[Q_{g}, \; Q_{b}] = +\; Q_b, \;\;\qquad  i\;[Q_{g}, \; Q_{ab}] = - \; Q_{ab},\;\; \qquad
 i\;[Q_{g}, \; Q_{ad}] = +\; Q_{ad},\;\;\qquad
 \nonumber\\ && \{Q_b,\; Q_d\} = Q_\omega \equiv -\; \{Q_{ad}, \; Q_{ab}\}, \quad
 [ Q_{\omega}\; Q_r ] = 0,
 \quad (r = b,\, ab,\, d,\, ad,\, g,\, {\omega} ).
\end{eqnarray}
The above algebra is exactly like the algebra satisfied by the de Rham  cohomological operators 
of differential geometry [7-12], namely;
\begin{eqnarray}
&&d^2 = 0,\quad \delta^2 = 0, \quad \{d, \delta\} = \Delta \equiv  (d +  \delta)^2, \quad
 [\Delta, d] = 0, \quad [\Delta, \delta] = 0,
\end{eqnarray}
where  $\delta = - * \,d \, *$ (with $\delta^2 = 0$) and $d= dx^\mu \partial_\mu$ (with $d^2 = 0$) are the (co-)exterior derivatives
and  $\Delta =  \{d,\, \delta\}$ is the Laplacian operator of differential geometry. In the above, 
the symbol ($*$) stands for the Hodge
duality operation on a given spacetime manifold. For the {\it even}
dimensional manifold, the relation  $\delta = - * \,d \, *$ is always true.

There is a simpler way to check the sanctity of the extended BRST algebra listed in (91) where we use
the well-known relationship between  the continuous symmetry transformations and their generators. For instance,
the above algebra can be obtained from the following transformations on the conserved charges, namely;
\begin{eqnarray}
&& s_r \, Q_r = i\, \{ Q_r, Q_r \} = 0 \;\; \qquad \Rightarrow \;\; 
 Q_r^2 = 0, \qquad\qquad\qquad (r = b, ab, d, ad), \nonumber\\
&& s_r \, Q_{ar} = i\, \{ Q_{ar}, Q_r \} = 0 \,\;\;\quad \Rightarrow \;\; \;
\{ Q_{ar}, Q_r \} = 0,\;\;\quad\qquad\qquad (r = b,  d), \nonumber\\
&& s_\omega \, Q_r = -\,i\, [ Q_r, Q_\omega ] = 0 \;\; \quad \;\Rightarrow \;\;\;\;
[ Q_\omega, Q_r ] = 0, \quad 
( r = b, ab, d, ad, g, \omega),\nonumber\\
&& s_g \, Q_r = -\,i\, [ Q_r, Q_g ] = +\, Q_r \;\,
 \Rightarrow \;\;
i\, [ Q_g, Q_r ] = + \,Q_r, \qquad (r = b, ad),\nonumber\\
&& s_b \, Q_{ad} = + i\; \{ Q_{ad}, Q_b \} = 0 \quad \Rightarrow \quad 
\{ Q_b, Q_{ad} \} = 0, \nonumber\\
&& s_d \, Q_{ab} = + i\; \{ Q_{ab}, Q_d \} = 0 \quad \Rightarrow \quad 
\{ Q_d, Q_{ab} \} = 0, \nonumber\\
&& s_g \, Q_r = -\,i\, [ Q_r, Q_g ] = -\, Q_r \;\; \Rightarrow\;\; 
i\, [ Q_g, Q_r ] = - \,Q_r, \qquad (r = d, ab), 
\end{eqnarray} 
where the l.h.s. can be calculated in a straightforward manner by exploiting the 
expressions for the {\it six} conserved charges and the corresponding continuous
symmetry transformations that have been mentioned in the main body of our text.

A comparison  between (91) and (92) demonstrates that $Q_\omega$ and $\Delta$ are 
the Casimir operators\footnote{The algebras (91) and (92) are {\it not} the Lie algebras.
Hence, the charge $Q_{\omega}$ and operator $\Delta$ are {\it not} 
the Casimir operators in the sense of such objects in the case of
Lie algebra.} for the above algebras in the sense that they 
absolutely commute with the rest of the operators.
 A close look at these algebras leads to the following clear-cut two-to-one mappings:
\begin{eqnarray}
&&(Q_{b}, Q_{ad}) \longrightarrow \,d,\quad (Q_{d}, Q_{ab}) \longrightarrow \,\delta, \quad
\{Q_{b}, \, Q_{d}\} = - \, \{Q_{ab},\, Q_{ad}\}   \longrightarrow  \,\Delta, 
\end{eqnarray}
between the conserved charges and the cohomological operators. Furthermore, we note that
we have the following beautiful relationship [2,3]: 
\begin{eqnarray}
&& s_{(a)d}\, \Psi = -\,*\,s_{(a)b}\,* \Psi, \qquad 
 (\Psi = A_\mu,\, \phi,\, \tilde \phi,\, C, \,\bar C,\, B,\, {\cal B}),
\end{eqnarray}
which provides the physical realization of the relationship 
(i.e. $\delta = -\,*\, d\,*$) between the (co-)exterior derivatives
$(\delta)d$   defined on an {\it even} dimensional spacetime manifold.
In the above equation (95), we observe that it is the interplay between the continuous symmetries
(i.e. $s_{(a)b},\;s_{(a)d}$) and the discrete symmetries (89) that provide the analogue of relationship 
$\delta = - *\, d\, *$. In fact, the latter [i.e. equation (89)] leads to the physical realization of 
the Hodge duality ($*$) operation of the differential geometry. Thus, we note  that the $(*)$
in (95) is nothing but the discrete symmetry transformations (89). 
The minus sign on the r.h.s of (95) is governed by
two successive operations of 
the discrete symmetry transformations (89)
on the generic field $\Psi$, namely; $* \, (*\, \Psi) = -\, \Psi$ (see, e.g. [33] for details).

One of the distinguishing features of the cohomological operators ($d,\,\delta,\, \Delta$) is the observation
that when they operate on a differential form of a specific degree, the consequences turn out to 
be completely different. For instance, when the (co-)exterior
derivatives operate on a form ($f_n$) of degree $n$, they (lower)raise the degree of the form 
by one (i.e. $\delta\,f_n \sim  f_{n-1}, \; d\, f_n \sim  f_{n+1}$). On the contrary, when $\Delta$ acts on a form
of degree $n$, it does {\it not} change the degree at all (i.e. $\Delta\,f_n \sim  f_n$). We have to capture these properties in
the language of the symmetry properties and conserved charges of our present modified version of
2D Proca theory so that we could establish precise analogy.

The above algebraic features could be also captured in the language of conserved charges. 
To this end in mind, let us define a state $|\psi \rangle_n$ in the quantum Hilbert space of states, as:
\begin{eqnarray}
i\, Q_g\, |\psi \rangle_n = n\, |\psi \rangle_n,
\end{eqnarray}
where the eigenvalue $n$ is the ghost number because $Q_g$ is the ghost charge [cf. (88)]. Due to
the algebra (91),  respected by the various charges, it can be readily checked that the
following relationships are true, namely;
\begin{eqnarray}
&& i\;Q_g\; Q_b\,|\psi\rangle_n = (n + 1) \;Q_b \;|\psi\rangle_n, \nonumber\\ &&
i\;Q_g\; Q_{ad}\,|\psi\rangle_n = (n + 1) \; Q_{ad} \;|\psi\rangle_n, \nonumber\\
&& i\;Q_g\; Q_d\,|\psi\rangle_n = (n - 1) \;Q_d \;|\psi\rangle_n, \nonumber\\ &&
i\;Q_g\; Q_{ab}\,|\psi\rangle_n = (n - 1) \; Q_{ab} \;|\psi\rangle_n, \nonumber\\
&& i\;Q_g\; Q_{\omega}\,|\psi\rangle_n = \;n \;Q_{\omega} \;|\psi\rangle_n.
\end{eqnarray}
Thus, we note that the ghost numbers for the  states $Q_b\,|\psi \rangle_n, \; Q_d\,|\psi \rangle_n$
and $Q_\omega\,|\psi \rangle_n$ are $(n+1), \; (n-1)$ and $n$, respectively. In exactly
similar fashion, the states  $Q_{ad}\,|\psi \rangle_n, \; Q_{ab}\,|\psi \rangle_n$ 
and $Q_\omega\,|\psi \rangle_n$ also carry the ghost numbers $(n+1), \; (n-1)$ and $n$, respectively. 
These properties are exactly like the consequences that ensue from the operations of the cohomological
operators $(d, \delta, \Delta )$ on a differential form of degree $n$ defined on a given manifold.

We conclude that, if the degree of a form is identified with the ghost number, then, the 
operation of ($d\,, \delta,\, \Delta$) on this given form is exactly like the operations of
the set ($Q_b,\;Q_d,\;Q_{\omega}$) and/or ($Q_{ad},\;Q_{ab},\;Q_{\omega}$) on the state 
with ghost number equal to
the degree of the form. Thus, the mappings (94) are {\it correct} as far as the algebraic structures
 of (91) and (92) are concerned and we have two-to-one mapping from the conserved charges of the theory 
to the de Rham cohomological
operators $(d, \delta, \Delta)$ of differential geometry.  A careful look at equations (90) and (91)
leads to the conclusion that the algebra (91) remains invariant under any number of
operations of discrete (duality) symmetry transformations (89). This establishes that our present
2D theory is a 
{\it perfect} model for the Hodge theory where the continuous symmetry transformations
(and corresponding generators) provide the physical realizations of the cohomological operators.
On the other hand, it is the discrete symmetry transformations of the theory that are the physical analogue of 
the Hodge duality $(*)$ operation of differential geometry. Finally, we observe that
the ghost number of a specific state in the quantum Hilbert space provides the physical
analogue of the degree of a form of differential geometry as far as its cohomological aspects are concerned.

\section{Conclusions}

In our present endeavor, we have applied the augmented version of superfield formalism to derive the
off-shell nilpotent (anti-)BRST and (anti-)co-BRST symmetry transformations for the modified version
of 2D Proca theory. We have exploited the theoretical strength of horizontality condition (HC) and
gauge invariant restriction (GIR) to derive the (anti-) BRST 
symmetries for {\it all} the fields of our present 2D theory.
In addition, we have made use of the dual-HC (DHC) and dual-GIR (DGIR) to obtain the complete set of 
(anti-)co-BRST symmetry transformations for {\it all} the fields of our present theory. The local gauge symmetry
transformations [cf. (4)] are the perfect ``classical'' version of the (anti-)BRST symmetries which
exist in any arbitrary dimension of spacetime. However, there is no such {\it perfect} ``classical'' 
analogue (see, e.g., subsection {\bf 4.1}) for the (anti-)co-BRST symmetries of our present theory. The latter 
symmetries exist only in specific dimensions of spacetime and they are always ``quantum'' in nature. 
For instance, for the
Abelian 1-form gauge theory, these ``quantum'' symmetries exist only in two dimensions of spacetime.

In our subsections {\bf 3.3} and {\bf 4.4}, we have expressed the (anti-)BRST and (anti-)co-BRST 
charges in various
forms due to our knowledge of the augmented  superfield approach to BRST formalism. In these subsections, 
we have been able to provide the meaning of their nilpotency and absolute
anticommutativity in the language of superfield formalism. We have  been 
{\it also} able to establish connections between the properties of 
nilpotency and absolute anticommutativity. In fact, it is the strength  
of the augmented superfield formalism that we have expressed the
(anti-)BRST and (anti-)co-BRST charges in a completely {\it novel} fashions (which have, hitherto,
not been pointed out in  literature). Thus, there are completely novel results in our subsections 
{\bf 3.3} and {\bf 4.4} as far as 
our present investigation on the superfield approach to BRST formalism is concerned.

In addition to the above results, there are applications of DHC and DGIR in 
deducing the full set of (anti-)co-BRST
symmetry transformations for {\it all} the fields of our present theory.
These derivations are {\it also} novel results. In particular, 
the application of DGIR, in the derivation of the (anti-)co-BRST 
symmetry transformations for the pseudo-scalar field ($\tilde \phi$), is a 
completely new result which has {\it not} been discussed 
in  literature. The symmetries of the theory enforce the pseudo-scalar field  
to have a negative kinetic term. Since this field  
is massive [i.e. $(\Box + m^2)\,\tilde\phi = 0$], it is a very good candidate  for the dark-matter [28,29].
We lay emphasis on the fact that the  Stuckelberg scalar field ($\phi$) has always a 
{\it positive} kinetic term  and, hence, it is an {\it ordinary} matter (due to $(\Box + m^2)\phi = 0)$.

In our investigation, we have provided physical realizations of the de Rham cohomological operators
in the language of the continuous symmetry transformations (and their corresponding charges). Further,
we have shown that a set of discrete symmetry transformations provide the physical analogue of the
Hodge duality $(*)$ operation of differential geometry. Ultimately, we have shown that, at the algebraic
level, the set of six conserved charges of our theory obey exactly the same algebra as that of the
de Rham cohomological operators of differential geometry. This algebra remains invariant [cf. (90)] 
under the discrete symmetry transformations (89) which are the analogue of Hodge theory ($*$) operation. 
The degree of a form finds its physical 
analogue as the ghost number of a state (in the quantum Hilbert space of states). Thus, our present 2D
modified version of Proca theory turns out to be a {\it perfect} model for the Hodge theory. The unique feature 
of our present theory is the co-existence of {\it mass}
and various kinds of {\it internal} symmetries {\it together} in a physically and mathematically meaningful manner.

It would be nice future endeavor to study the above kind of possibilities in the cases of
3D and 4D massive gauge theories [34,35] where the gauge invariance and  mass 
would co-exist together. In other words, we
would like to study whether Stueckelberg's type of technique would be able to modify the above theories in such a way
that they could also become {\it massive} field theoretic  models for the Hodge theory. We speculate that such kind
of situation will exist and these models will provide candidates for the {\it dark} matter in more physical 3D and 4D
of spacetime (analogous to the massive pseudo-scalar $\tilde \phi$ of our present 2D theory). 
Our speculation is based on the fact that we have already shown that the 4D free Abelian 2-form gauge theory is
a model for the Hodge theory where a {\it massless} 
pseudo-scalar field does exist with a {\it negative} kinetic term
(see, e.g. [17,18] for details). We are currently intensively
involved with such kind of problems and,  we shall be able to report about our progress in our
future publications.\\

\noindent
{\bf Acknowledgements}:
Two of us (AS and SK) would like to gratefully acknowledge  the financial support
from CSIR and UGC, Government of India, New Delhi, under their respective  SRF-schemes.\\

\vskip .5cm

\noindent
{\bf{\large{~~~~~~~~~~~~~~~~Appendix A: On the verification of (46)} }}
\vskip .5cm

\noindent
Here we compute equation (46) step-by-step which is nothing but the expression for ($- \star  \tilde d \star 
\tilde A^{(1)}$). Taking the expression for $\tilde A^{(1)}$ (from 6) and applying
a single ($\star$) on it, we obtain the following on a (2, 2)-dimensional supermanifold (see, e.g. [31] for details)
\[\star \,\tilde A^{(1)} = \varepsilon^{\mu\nu}\, 
(dx_\nu\wedge d\theta \wedge d\bar\theta) \, B_\mu(x, \theta, \bar\theta) 
+ \frac{1}{2!}\,\varepsilon_{\mu\nu}\,(dx^\mu\wedge d x^\nu \wedge d\bar\theta) \,
 {\bar F}(x, \theta, \bar\theta)\]
\[+ \frac{1}{2!}\,\varepsilon_{\mu\nu}\,(dx^\mu\wedge d x^\nu \wedge d \theta) \,
 { F}(x, \theta, \bar\theta), ~~~~~~~~~~~~~~~~~~~~~~~~~~~~~~~~~~\eqno(A.1)\]  
which is nothing but a super 3-form on the above supermanifold. In the above computation, 
we have used the following relationship on the given (2, 2)-dimensional supermanifold (see, e.g. [31] for details):
\[\star\, (dx^\mu) = \varepsilon^{\mu\nu}\, 
(dx_\nu\wedge d\theta \wedge d\bar\theta),~~~\]  
\[\star\; (d\theta) = \frac{1}{2!}\,\varepsilon_{\mu\nu}\,(dx^\mu\wedge d x^\nu \wedge d\bar\theta), \]
\[\star \;(d \bar\theta) = \frac{1}{2!}\,\varepsilon_{\mu\nu}\,(dx^\mu\wedge d x^\nu \wedge d \theta).\eqno(A.2)\] 
Now we obtain a super 4-form on the (2, 2)-dimensional supermanifold by applying a $\tilde d$ on (A.1).
This is given by the following expression:
\[\tilde d\star \tilde A^{(1)} = \varepsilon^{\mu\nu}\, 
(d x_\lambda \wedge dx_\nu \wedge    d\theta \wedge d\bar\theta) \, 
\partial^\lambda B_\mu(x, \theta, \bar\theta) ~~~~~~~~~~~ \]
\[~~~ +\, \frac{1}{2!}\,\varepsilon_{\mu\nu}\, (dx^\lambda\wedge d x^\mu \wedge  d x^\nu \wedge d\bar\theta)\, 
\partial_\lambda \,{\bar F}(x, \theta, \bar\theta)\]
\[~~~ +\, \frac{1}{2!}\,\varepsilon_{\mu\nu}\, (dx^\lambda\wedge d x^\mu \wedge  d x^\nu \wedge d\theta)\, 
\partial_\lambda \,{ F}(x, \theta, \bar\theta)\]  
\[+ \,\varepsilon^{\mu\nu}\, (d\theta\wedge d x_\nu \wedge  d\theta \wedge d\bar\theta)
 \, \partial_{\theta} B_\mu(x, \theta, \bar\theta) ~~~\]
\[- \, \frac{1}{2!}\,\varepsilon_{\mu\nu}\, (d \theta \wedge  d x^\mu \wedge  d x^\nu \wedge d\bar\theta)\, 
\partial_\theta \,{ \bar F}(x, \theta, \bar\theta) \]  
\[-\, \frac{1}{2!}\,\varepsilon_{\mu\nu}\, (d \theta \wedge  d x^\mu \wedge  d x^\nu \wedge d\bar\theta)\, 
\partial_\theta \,{  F}(x, \theta, \bar\theta)\] 
\[+\,  \varepsilon^{\mu\nu}\, 
(d\bar\theta\wedge d x_\nu \wedge  d\theta \wedge d\bar\theta) 
\, \partial_{\bar\theta} B_\mu(x, \theta, \bar\theta) \]
\[- \, \frac{1}{2!}\,\varepsilon_{\mu\nu}\, (d \bar\theta \wedge  d x^\mu \wedge  d x^\nu \wedge d\bar\theta)\, 
\partial_{\bar\theta} \,{ \bar F}(x, \theta, \bar\theta) \]
\[-\, \frac{1}{2!}\,\varepsilon_{\mu\nu}\, (d \bar\theta \wedge  d x^\mu \wedge dx^\nu \wedge d\theta)\, 
\partial_{\bar\theta} \,{F}(x, \theta, \bar\theta). \eqno(A.3)\] 
It is clear, from the above, that the second and third terms would be zero because
there are wedge-products which contains {\it three} spacetime differentials (which is {\it not}
allowed on a (2, 2)-dimensional supermanifold). Furthermore, fourth and seventh terms would be zero because
the wedge product with {\it three} Grassmannian differentials are {\it not}
allowed on a (2, 2)-dimensional supermanifold. Hence, we have the existing super 4-form as:
\[\tilde d\star \tilde A^{(1)} = \varepsilon^{\mu\nu}\, 
(d x_\lambda \wedge dx_\nu\wedge   d\theta \wedge d\bar\theta) \, 
\partial^\lambda B_\mu(x, \theta, \bar\theta) ~~~~~~~~~~~~\]
\[- \frac{1}{2!}\,\varepsilon_{\mu\nu}\, (d \theta \wedge  d x^\mu \wedge  d x^\nu \wedge d\bar\theta)\, 
\partial_\theta \,{ \bar F}(x, \theta, \bar\theta) \] 
\[- \frac{1}{2!}\,\varepsilon_{\mu\nu}\, (d x^\mu \wedge  d x^\nu  \wedge d \theta \wedge d\bar\theta)\, 
\partial_\theta \,{  F}(x, \theta, \bar\theta) \] 
\[- \frac{1}{2!}\,\varepsilon_{\mu\nu}\, (d \bar\theta \wedge  d x^\mu \wedge  d x^\nu \wedge d\bar\theta)\, 
\partial_{\bar\theta} \,{ \bar F}(x, \theta, \bar\theta) \]
\[-\,\frac{1}{2!}\,\varepsilon_{\mu\nu}\, (d \bar\theta \wedge  d x^\mu \wedge  d x^\nu \wedge d\theta)\, 
\partial_{\bar\theta} \,{  F}(x, \theta, \bar\theta). \eqno(A.4)\] 
where we have taken into account the following rules:
\[(dx^\mu \wedge d \theta) =   -\,(d \theta \wedge dx^\mu), \qquad  \qquad
(d\theta \wedge d \bar\theta) = (d \bar\theta \wedge d \theta). \eqno(A.5)\]
Taking a [$- (\star)$] on (A.4), we obtain:
\[ -\,\star\, \tilde d\star \tilde A^{(1)} = - \,\varepsilon^{\mu\nu}\,\varepsilon_{\lambda\nu} \partial^\lambda B_\mu
- \partial_\theta \bar{\cal F } - S^{\bar\theta \bar\theta} \partial_{\bar\theta} \bar{\cal F } 
- S^{\theta \theta} \partial_{\theta} {\cal F } -\partial_{\bar\theta} {\cal F }, \eqno(A.6)\]
where we have used the following inputs (see, e.g. [31]) 
\[ \star\, (dx_\lambda \wedge dx_\nu \wedge d\theta \wedge d\bar\theta) = \varepsilon_{\lambda\nu}, \qquad\qquad
\star\, (dx^\mu\wedge dx^\nu \wedge d\theta \wedge d\bar\theta) = \varepsilon^{\mu\nu}, ~~~~~~~~~~ \]
\[\star\, (dx^\mu \wedge dx^\nu \wedge d\theta \wedge d\theta) = \varepsilon^{\mu\nu}\, S^{\theta \theta}, \qquad
\star\, (dx^\mu \wedge dx^\nu \wedge d\bar\theta \wedge d\bar\theta) = \varepsilon^{\mu\nu}\, 
S^{\bar\theta \bar\theta}. \eqno(A.7)\]
Thus, we finally obtain the 
explicit expression for ($-\,\star\, \tilde d\star \tilde A^{(1)}$) as follows:
\[-\,\star\, \tilde d\star \tilde A^{(1)} = (\partial \cdot {\cal B}) - (\partial_\theta \bar F
+ \partial_{\bar\theta} {\cal F})- S^{\theta\theta} (\partial_\theta F)  
- S^{\bar\theta \bar\theta} (\partial_{\bar\theta} \bar F), \eqno(A.8)\]
which has been mentioned in our equation (46).\\

\vskip .5cm

\noindent
{\bf{\large{~~~~~~~~~~~~~~~~~~Appendix B: On the verification of (49)}}}
\vskip .5cm

\noindent
By exploiting the basic ideas behind the augmented superfield formulation, 
we demonstrate here that the choices made in (49)
are {\it exact}. Towards this goal in mind, we note that the following (anti-)co-BRST invariant quantity:
\[s_{(a)d} \left[\varepsilon^{\mu\nu} (\partial_\mu {\cal B}) A_\nu -  i\, \partial_\mu {\bar C}
 \partial^\mu C \right] = 0, \eqno(B.1)\]
should remain independent of the ``soul" coordinates $\theta$ and $\bar\theta$
when it is generalized onto the (1, 1)-dimensional (anti-)chiral super-submanifolds. 
This is physically allowed and it can be readily utilized within the framework
of the augmented superfield formalism. In other words, the following 
equality:
\[ \varepsilon^{\mu\nu} \,(\partial_\mu {\cal B}(x)) \,{\cal B}_\nu (x, \theta, \bar\theta) 
 - i\, \partial_\mu \bar{ F}^{(dh)}(x, \theta, \bar\theta)\,
\partial^\mu {F}^{(dh)}(x, \theta, \bar\theta) \]
\[= \varepsilon^{\mu\nu}\, (\partial_\mu {\cal B} (x))\, A_\nu (x) - i\, \partial_\mu\, {\bar C}(x)\,
 \partial^\mu C (x), \eqno(B.2)\]
should hold good as far as the (super)fields of our present theory are concerned. 
Plugging in the super-expansions from (7) and (50)
for ${\cal B}_\mu (x, \theta, \bar\theta),\;\,{ F}^{(dh)}(x, \theta, \bar\theta)$ and 
$\bar{ F}^{(dh)}(x, \theta, \bar\theta)$,
we obtain the following relationships:
\[ \varepsilon^{\mu\nu} \left(\partial_\mu {\cal B} (x)\right)\, {\bar R}_\nu (x) 
+ \partial_\mu C(x)\, \partial^\mu {\cal B}(x) = 0, \]
\[ \varepsilon^{\mu\nu} (\partial_\mu {\cal B} (x))\,  R_\nu (x) 
+ \partial_\mu {\bar C}(x)\, \partial^\mu {\cal B}(x) = 0, \]
\[ \varepsilon^{\mu\nu} (\partial_\mu {\cal B} (x))\, S_\nu (x) 
+ \partial_\mu {\cal B}(x)\, \partial^\mu {\cal B}(x) = 0, \eqno(B.3)\]
which are obtained when we set equal to zero the coefficients of
 $\theta$ and $\bar\theta$ and $\theta\,\bar\theta$ in the
equality (B.2). From (B.3), it is clear that we obtain:
\[ {\bar R}_\mu = -\varepsilon_{\mu\nu}\, \partial^\nu C, \qquad R_\mu 
= -\varepsilon_{\mu\nu}\, \partial^\nu {\bar C},\qquad 
 S_\mu = \varepsilon_{\mu\nu}\, \partial^\nu {\cal B}, \eqno(B.4)\]
The substitution of these values into (7) leads to the derivation of the expansions ${\cal B}_\mu^{(dh)}(x, \theta, \bar\theta)$.
It is worth pointing out that the expansions for ${ F}^{(dh)}(x, \theta, \bar\theta)$ and $\bar { F}^{(dh)}(x, \theta, \bar\theta)$
have been obtained due to 
the dual-HC (given in (45)). This demonstrates that, for our 2D theory, the choices made in (49) can be computed 
{\it exactly}
in a precise manner.




%
%

\end{document}